%% file: main.tex
\documentclass[preprint,12pt]{elsarticle}
\usepackage{amsmath,amssymb,booktabs,microtype,array}
\usepackage[section]{placeins}
\usepackage[hidelinks]{hyperref}
\journal{Journal of Computational Physics}
\newcommand{\paperfig}[3]{\begin{figure}[htbp]\centering\includegraphics[width=\linewidth,height=.77\textheight,keepaspectratio]{#1}\caption{#2}\label{#3}\end{figure}}
\begin{document}
\begin{frontmatter}
\title{A centered finite-volume method with explicit pressure relaxation for weakly compressible single- and two-phase flows}
\author[tokyo]{Shintaro Matsushita\corref{cor1}}
\ead{matsushita.s.90b7@m.isct.ac.jp}
\author[cmfr,gt]{Suhas S. Jain}
\author[gt,ibb]{Andrei G. Fedorov}
\author[tokyo]{Tetsuya~Suekane}
\cortext[cor1]{Corresponding author.}
\affiliation[tokyo]{organization={Department of Mechanical Engineering, Institute of Science Tokyo},city={Tokyo},postcode={152-8550},country={Japan}}
\affiliation[cmfr]{organization={Center for Multiphase flow Research, Georgia Institute of Technology},city={Atlanta},state={GA},postcode={30332},country={USA}}
\affiliation[gt]{organization={George W. Woodruff School of Mechanical Engineering, Georgia Institute of Technology},city={Atlanta},state={GA},postcode={30332},country={USA}}
\affiliation[ibb]{organization={Parker H. Petit Institute for Bioengineering and Bioscience, Georgia Institute of Technology},city={Atlanta},state={GA},postcode={30332},country={USA}}
\begin{abstract}
We present a non-dissipative centered finite-volume method for weakly compressible single- and two-phase flows. The method couples kinetic-energy-preserving advection to conservative diffuse-interface regularization and explicit pressure relaxation on structured collocated grids. To control excessive phase-mass outflow when neighboring cells have different intrinsic densities, we construct phase-mass fluxes using harmonic face densities. This connects phase-volume and phase-mass transport without altering the advective kinetic-energy identity, yielding a sufficient positivity condition and an inverse-density transport identity at finite velocity divergence. To retain the benefit of low-dissipation transport in the coupled calculation, a pressure-increment formulation incorporates the existing pressure gradient in the predictor. Compatible cell and face operators give a conditional energy estimate for additional pressure iterations, while fully discrete budgets separate advection, predictor pressure work, and the final correction. In inviscid homogeneous isotropic turbulence on $128^3$ cells, the relative kinetic-energy change remains below $1.2\times10^{-7}$ in magnitude at $t^*=8$ in both single-phase and density-ratio-$1000$ two-phase calculations. Matched upwind calculations retain only $3.60\%$ and $30.72\%$ of their initial energies, respectively. Shear layers, cavities, rising bubbles, and dam breaks demonstrate applicability to incompressible-regime flows. In a three-dimensional dam break, harmonic averaging retains positive phase masses where arithmetic averaging produces negative values. With centered advection, a prescribed subgrid-scale stress supplies the principal dissipation. The local pressure algorithm and collocated finite-volume form support future large-scale multiphase large-eddy simulation on GPUs with adaptive mesh refinement.
\end{abstract}
\end{frontmatter}
\input{introduction}
\input{methods}
\input{properties}
\input{verification}
\input{hit}
\input{single_phase}
\input{two_phase}
\input{three_dimensional}
\input{discussion}
\input{appendices}
\input{backmatter}
\begingroup\small
\setlength{\bibsep}{5pt}
\bibliographystyle{elsarticle-num}
\bibliography{references}
\endgroup
\end{document}

%% file: introduction.tex
\section{Introduction}\label{sec:intro}

High-resolution simulation of low-Mach-number multiphase flow must resolve vortical motion and deforming interfaces across widely separated scales. Two numerical requirements are particularly important: momentum transport must retain the resolved structures, and pressure computation must scale to the large cell counts needed to represent them. Upwind numerical viscosity can damp those structures before molecular viscosity or a prescribed subgrid-scale (SGS) model acts. A computationally scalable pressure algorithm alone therefore does not provide a low-dissipation multiphase solver. This work addresses both requirements through a centered weakly compressible finite-volume formulation.

In incompressible projection methods, a pressure Poisson equation enforces the velocity-divergence constraint. Efficient solvers and pressure splittings have substantially improved this approach. Dodd and Ferrante \cite{DoddFerrante2014}, for example, reduced a variable-coefficient two-fluid pressure problem to a constant-coefficient solve. The elliptic problem nevertheless couples the computational domain through its convergence process. Explicit weakly compressible methods instead evolve pressure with a finite artificial acoustic speed, replacing that solve with local updates. Although the acoustic scale restricts the time step, the compact stencil and predictable communication pattern are well suited to parallel computation on GPUs and adaptive meshes.

The use of weak compressibility for nominally incompressible flow is well established in lattice Boltzmann methods (LBMs), which retain a finite acoustic speed and approach the incompressible regime at low Mach number. Multiphase applications include three-dimensional dam breaks and liquid-jet breakup \cite{SitompulAoki2019}. Holzer et al.\ \cite{Holzer2021} demonstrated simulations of several hundred resolved bubbles on up to 2048 GPUs. These results demonstrate the suitability of local weakly compressible algorithms for large-scale multiphase simulation.

In our previous work, we developed a weakly compressible diffuse-interface method for low-Mach-number two-phase flows \cite{Matsushita2019} and extended it to interface-adapted adaptive mesh refinement (AMR) \cite{Matsushita2021}. An expanding liquid film of approximately $0.1$ mm thickness was resolved using about $0.9$ million cells, with a finest resolution equivalent to an $8192\times8192$ uniform mesh. A wet-floor dam-break performance test achieved weak-scaling efficiencies of approximately $81\%$ for the complete AMR computation and $95\%$ for the fluid update on 128 GPUs of the FLOW supercomputer. Thus weak compressibility has already enabled high-resolution multiphase simulations on adaptive meshes with demonstrated many-GPU scalability. Evolving pressure projection (EPP) suppresses acoustic oscillations through local pressure relaxation \cite{Yang2021}, and has also been formulated on collocated grids with momentum conservation \cite{YangCollocated}. These developments provide the pressure and parallel-computing basis for the present study.

For this computational approach to support large-eddy simulation (LES) with explicit SGS models, the role of upwind dissipation must be addressed. Recent weakly compressible methods include the HLLC formulation of Melvin and Mandal \cite{MelvinMandal2025} and their path-conservative extension with surface tension \cite{MelvinMandal2026}. Upwind fluxes promote robustness, but they also dissipate resolved kinetic energy. When this numerical dissipation supplies a significant part of the unresolved-scale loss, the discretization acts as an implicit SGS closure alongside any explicit model. Numerical error and SGS modeling can therefore interact \cite{Ghosal1996}. Eliminating the upwind advection sink allows molecular and SGS stresses to act on the retained velocity gradients and makes their contributions to energy transfer directly assessable.

Kinetic-energy-preserving spatial discretizations provide a basis for this objective. Morinishi et al.\ \cite{Morinishi1998} established conservation requirements for incompressible finite-difference schemes, and Morinishi \cite{Morinishi2010} developed skew-symmetric formulations for variable-density low-Mach flow. For diffuse-interface two-phase flow, Mirjalili and Mani \cite{MirjaliliMani2021} constructed consistent mass and momentum transport on a staggered grid to preserve kinetic energy. Manzanero et al.\ \cite{Manzanero2020} developed an energy-consistent artificial-compressibility discontinuous Galerkin method coupled to Cahn--Hilliard dynamics. These methods show that nondissipative transport requires the discrete mass and momentum equations to satisfy the kinetic-energy balance together. A collocated finite-volume realization is especially attractive for the present application because all momentum components share one cell hierarchy and conservative face fluxes can be used in AMR.

Kinetic-energy- and entropy-preserving (KEEP) schemes construct centered fluxes using these discrete relations for compressible flow \cite{Kuya2018}. Jain and Moin \cite{JainMoin2022} extended the formulation to compressible two-phase flow, including mass transport due to interface regularization in the kinetic-energy balance. For the interface, Jain's accurate conservative diffuse-interface (ACDI) method \cite{JainACDI} combines centered transport with balanced diffusion and sharpening. Hwang and Jain \cite{HwangJain2024} subsequently developed an incompressible ACDI method on unstructured collocated grids using a pressure Poisson solve. These advances show that large-density-ratio two-phase transport can be formulated with centered discretization while accounting for the interface-induced mass flux. Explicit weakly compressible pressure evolution offers a way to combine this low-dissipation transport with the local computational structure of a structured-grid solver.

The target low-Mach flows involve vortical motion and interface deformation without shocks, making centered transport particularly attractive. Interface-profile control is also useful in this regime: weak compression can deform a diffuse-interface profile, and our previous study \cite{Matsushita2021} discussed its relaxation by phase-field diffusion. ACDI provides this diffusion--sharpening balance as the interface is transported. It can therefore be coupled to EPP with distinct roles: ACDI maintains the interface profile, and EPP suppresses acoustic pressure oscillations.

A central difficulty in extending centered transport to this weakly compressible setting is the construction of the phase-mass flux. Phase volume and phase mass evolve independently at finite velocity divergence, so the intrinsic density, given by phase mass divided by phase volume, also evolves. Large relative contrasts can develop between neighboring cells where a phase occupies a small volume fraction. An arithmetic face density can then carry excessive mass out of the cell with lower intrinsic density. Sharing the flux conserves total phase mass, but does not by itself prevent a negative local phase mass. The phase-mass flux must therefore control local outflow without sacrificing the kinetic-energy balance of centered momentum transport.

We address this difficulty by applying a harmonic mean to the intrinsic density of each phase on the two sides of a face. The resulting face density is bounded by twice the intrinsic density in either cell. This gives a local bound on outgoing mass and a sufficient positivity condition in terms of the actual phase-volume outflow, including interface regularization. At the same time, the centered velocity average preserves the advective kinetic-energy cancellation with this shared mass flux. Harmonic averaging also cancels internal-face transport in an inverse-density functional, leaving only the effect of finite velocity divergence. Harmonic means have been used in pressure-equilibrium-preserving compressible fluxes \cite{DeMicheleCoppola2024}; here they connect independently evolved phase volumes and masses and reconcile positive mass transport with centered momentum advection. The resulting outflow bound does not require an exactly divergence-free velocity, making it directly applicable to weakly compressible two-phase flow.

A second issue is how pressure relaxation is coupled to nondissipative transport. In the collocated EPP formulation of Yang and Aoki \cite{YangCollocated}, pressure is relaxed using face velocities and the resulting pressure force is interpolated to the cell velocities. Here, trial face velocities are constructed from the pressure-corrected cell momentum, so that the pressure gradient and velocity divergence act on a common discrete kinetic energy. We also include the existing pressure gradient throughout the momentum predictor and subsequently correct the momentum with a pressure increment. This temporal connection is important when advection itself supplies no energy sink: an unnecessarily dissipative pressure split would obscure the benefit of centered transport. The resulting local explicit relaxation admits a conditional energy estimate, and its complete pressure work can be measured independently of advection.

The proposed structured collocated finite-volume method uses centered spatial discretization for advection, viscous stress, pressure, interface regularization, and surface tension. Harmonic phase-mass transport provides an outflow bound while retaining the advective kinetic-energy identity; pressure-increment EPP connects this transport to local explicit pressure relaxation. Their combined construction allows large interface deformation to be computed without upwind momentum dissipation. We establish the discrete transport and pressure properties and use fully discrete energy budgets to determine how they carry through to the coupled calculation.

The assessment compares the proposed centered method with an upwind variant constructed solely for measuring the effect of momentum-advection dissipation. Phase transport, EPP, and physical models are held common, and fully discrete energy budgets separate advective, molecular-viscous, SGS, and pressure contributions. Inviscid homogeneous isotropic turbulence (HIT) in single- and two-phase flow quantifies the energy retained by the complete algorithm and the effect of pressure coupling. Paired changes of momentum reconstruction and intrinsic-density averaging distinguish their respective effects on energy and phase-mass transport. Further benchmarks evaluate vortical transport, agreement with incompressible reference solutions, and capillary and buoyant motion. Three-dimensional dam breaks at density ratio $1000$ then demonstrate the improvement in phase-mass transport and quantify the competition between upwind dissipation and an explicit SGS closure. These comparisons establish how low-dissipation transport changes the action of the explicit SGS model within the weakly compressible solver. Its common cell placement and conservative face fluxes also facilitate future integration into our GPU AMR framework, which already employs finite-volume phase-field transport \cite{Matsushita2021}.

%% file: methods.tex
\section{Governing model and numerical method}\label{sec:method}

\subsection{Independent phase variables and weakly compressible model}\label{sec:governing}

We consider isothermal, immiscible Newtonian fluids without phase change. A common velocity $\boldsymbol u$ and pressure $p$ are used. The liquid volume fraction $\phi$, partial mass densities $m_1,m_2$, momentum density $\boldsymbol M$, and pressure are independent cell variables. For phases $\ell=1,2$, define

\begin{equation}\label{eq:variables}
\alpha_1=\phi,\quad \alpha_2=1-\phi,\quad
\rho=m_1+m_2,\quad \boldsymbol u=\frac{\boldsymbol M}{\rho},\quad
r_\ell=\frac{m_\ell}{\alpha_\ell}.
\end{equation}

Thus $m_\ell$ is mass per mixture volume, whereas $r_\ell$ is the intrinsic density of phase $\ell$. The fixed reference densities $\rho_\ell^0$ initialize $m_\ell=\rho_\ell^0\alpha_\ell$; this algebraic relation is not imposed during the evolution. This distinction allows phase mass to remain conservative while the velocity has the small, finite divergence admitted by the pressure model.

\begin{equation}\label{eq:phase_model}
\begin{aligned}
\partial_t\alpha_\ell+\nabla\cdot\boldsymbol j_\ell
 &=\alpha_\ell\nabla\cdot\boldsymbol u,
 &\boldsymbol j_\ell&=\alpha_\ell\boldsymbol u-\boldsymbol a_\ell,\\
\partial_t m_\ell+\nabla\cdot\boldsymbol f_\ell&=0,
 &\boldsymbol f_\ell&=r_\ell\boldsymbol j_\ell,\\
\boldsymbol a_2&=-\boldsymbol a_1,
 &\boldsymbol f&=\boldsymbol f_1+\boldsymbol f_2.
\end{aligned}
\end{equation}

\begin{equation}\label{eq:momentum_pressure_model}
\partial_t\boldsymbol M+\nabla\cdot(\boldsymbol u\otimes\boldsymbol f)
=-\nabla p+\nabla\cdot\boldsymbol\tau+\rho\boldsymbol g+\boldsymbol f_\sigma,
\qquad \partial_t p=-\rho c^2\nabla\cdot\boldsymbol u.
\end{equation}

Here $\boldsymbol a_1$ is the interface-regularization volume flux, $c$ is the artificial acoustic speed, and $\boldsymbol f_\sigma$ is the surface-tension force per volume. Summing the phase-mass equations gives $\partial_t\rho+\nabla\cdot\boldsymbol f=0$, with exactly the mass flux entering momentum transport. Summing the volume-fraction equations preserves $\alpha_1+\alpha_2=1$. The volume-fraction source arises when material advection is written in flux form. At finite divergence, the formulation conserves phase mass while allowing phase volume to change.

The pressure equation closes the isothermal weakly compressible model. Scaling pressure by $\rho_0U_{\mathrm{ref}}^2$, with reference density $\rho_0$, gives a coefficient proportional to $M_{\mathrm{ref}}^{-2}$, where $M_{\mathrm{ref}}=U_{\mathrm{ref}}/c$. With bounded pressure evolution and positive density, the small-$M_{\mathrm{ref}}$ balance gives a correspondingly small velocity divergence. A finite artificial Mach number and EPP relaxation are used to compute flows in this regime.

\begin{equation}\label{eq:stress}
\tau_{ab}=\mu\left(\partial_bu_a+\partial_au_b
-\frac23\delta_{ab}\nabla\cdot\boldsymbol u\right),\qquad
\mu=\mu_1\phi+\mu_2(1-\phi).
\end{equation}

The inputs $\mu_\ell$ are dynamic viscosities. The coefficient $2/3$ is retained in two-dimensional calculations, corresponding to a three-dimensional constitutive law with zero out-of-plane velocity.

\subsection{Finite-bound ACDI interface regularization}\label{sec:acdi}

For two-phase calculations, we use the finite-bound ACDI regularization of Hatashita and Jain \cite{HatashitaJain2026}. The pure-phase limits are represented by $\delta=10^{-8}$ and $1-\delta$. Set $s=1-2\delta$ and initialize $\phi=\delta+s\chi$, with $\chi\in[0,1]$. The ACDI auxiliary distance and normal are

\begin{equation}\label{eq:acdi_continuum}
\begin{aligned}
\psi&=\epsilon\log\frac{\phi-\delta}{1-\phi-\delta},\qquad
\boldsymbol n=\frac{\nabla\psi}{|\nabla\psi|},\\
\boldsymbol a_1&=\Gamma\left[\epsilon s\nabla\phi
-\frac{s^2}{4}\operatorname{sech}^2\!\left(\frac{\psi}{2\epsilon}\right)\boldsymbol n\right].
\end{aligned}
\end{equation}

The width $\epsilon$ and regularization speed $\Gamma$ set the diffusion--sharpening balance. A signed-distance profile $\phi=\delta+\tfrac{s}{2}[1+\tanh(\psi/(2\epsilon))]$ makes these terms balance when $|\nabla\psi|=1$. The auxiliary variable is evaluated from the current phase fraction and is not transported as another unknown.

The logarithm is evaluated using temporary arguments restricted to $[0,s]$, with $\eta=10^{-100}$ added to numerator and denominator. This defines the logarithm at the finite pure-phase limits without modifying the stored fields. The normal is set to zero where its gradient vanishes. In the limit $\delta\to0$, Eq.~\eqref{eq:acdi_continuum} recovers the standard ACDI profile relation \cite{JainACDI}.

\subsection{Shared face fluxes on a collocated finite-volume grid}\label{sec:flux}

The method uses centered spatial discretization throughout: symmetric face interpolation for transport, centered differences for viscous and pressure derivatives, and centered ACDI and capillary forces. All primary variables are stored at Cartesian cell centers. Let $V_i$ be a cell volume, $S_f$ a face area, and $s_{if}=\pm1$ the outward orientation of face $f$ relative to cell $i$. A single face flux is shared by adjacent cells. The finite-volume divergence and arithmetic face average are

\begin{equation}\label{eq:operators}
(Dq)_i=\frac1{V_i}\sum_{f\in\partial i}s_{if}S_fq_f,
\qquad \overline q_f=\frac{q_L+q_R}{2}.
\end{equation}

For a face with unit normal $\boldsymbol e_f$ and spacing $h$, the centered ACDI regularization is

\begin{equation}\label{eq:acdi_face}
A_{1,f}=\Gamma\left[\epsilon s\frac{\phi_R-\phi_L}{h}
-\frac{s^2}{4}\operatorname{sech}^2\!\left(\frac{\psi_L+\psi_R}{4\epsilon}\right)
\overline{\boldsymbol n}_f\cdot\boldsymbol e_f\right],\qquad A_{2,f}=-A_{1,f}.
\end{equation}

Cell normals use centered gradients of $\psi$, and their arithmetic average is used at a face. The normal advecting velocity is denoted by $U_f$. The volume and mass fluxes are

\begin{equation}\label{eq:phase_fluxes}
F_{\alpha_\ell,f}=\overline{\alpha_\ell}_fU_f-A_{\ell,f},\qquad
F_{m_\ell,f}=r^H_{\ell,f}F_{\alpha_\ell,f},\qquad
F_{m,f}=F_{m_1,f}+F_{m_2,f}.
\end{equation}

\begin{equation}\label{eq:harmonic}
r^H_{\ell,f}=\frac{2r_{\ell,L}r_{\ell,R}}{r_{\ell,L}+r_{\ell,R}}
=\frac{2}{\alpha_{\ell,L}/m_{\ell,L}+\alpha_{\ell,R}/m_{\ell,R}}.
\end{equation}

Here $r_{\ell,L}$ and $r_{\ell,R}$ are the intrinsic densities of the same phase $\ell$ in the two cells adjacent to face $f$. Equation~\eqref{eq:harmonic} is symmetric under interchange of $L$ and $R$ and thus retains a centered phase-mass flux. Harmonic density means have also been used in pressure-equilibrium-preserving compressible Euler schemes \cite{DeMicheleCoppola2024}. In the present formulation, the mean relates the independently evolved phase volume and phase mass through Eq.~\eqref{eq:phase_fluxes}, including ACDI transport. Its bound on outgoing mass and inverse-density transport identity are derived in Section~\ref{sec:h_properties}.

\begin{equation}\label{eq:semidiscrete_phase}
\dot\alpha_{\ell,i}=-(DF_{\alpha_\ell})_i+\alpha_{\ell,i}\theta_i,
\qquad \dot m_{\ell,i}=-(DF_{m_\ell})_i,\qquad
\theta_i=(DU)_i.
\end{equation}

Only $\alpha_1$ is stored; $\alpha_2=1-\alpha_1$ supplies the complementary equation. The momentum-advection flux and the non-pressure part of the transport predictor are

\begin{equation}\label{eq:momentum_flux}
\boldsymbol F_{M,f}=F_{m,f}\overline{\boldsymbol u}_f,
\qquad \boldsymbol R_M=-D\boldsymbol F_M+D\boldsymbol\tau+\rho\boldsymbol g.
\end{equation}

The same actual mass flux is used in the density and momentum updates. Multiplication by the symmetric cell-velocity average makes the advective kinetic-energy cancellation independent of the detailed phase-volume flux. The transport identity therefore includes the mass redistribution caused by interface regularization.

\subsection{Viscous and capillary forces}\label{sec:forces}

Variable viscosity is placed consistently with the derivative being approximated. At a face normal to direction $d$, let $\delta_d u_a=(u_{a,R}-u_{a,L})/h$ and let $D_a^c$ denote the centered cell derivative. The stress components used in the face divergence are

\begin{equation}\label{eq:viscous_face}
\begin{aligned}
\tau_{ad,f}&=\overline\mu_f\,\delta_d u_a+
\overline{\mu D_a^c u_d}_f,&&a\ne d,\\
\tau_{dd,f}&=2\overline\mu_f\,\delta_d u_d
-\frac23\left(\overline\mu_f\,\delta_d u_d+
\sum_{b\ne d}\overline{\mu D_b^c u_b}_f\right).
\end{aligned}
\end{equation}

Normal derivatives use the arithmetic face viscosity. Transverse derivatives use the arithmetic average of the viscosity-weighted cell derivative. For constant viscosity these reduce to the usual centered stresses. The placement of $\mu$ inside the transverse average is retained for both molecular and SGS stresses. It yields a symmetric dissipative operator in the periodic setting considered in \ref{app:viscous}.

Surface tension is represented by a continuum surface force (CSF) \cite{Brackbill1992}. For curvature, the gradient of $\psi$ is first evaluated at each face: its normal component uses the adjacent-cell difference, and its tangential components use averages of centered cell derivatives. Normalizing this face gradient gives $\boldsymbol n_f^\sigma$, and $\kappa_i=-[D(\boldsymbol n_f^\sigma\cdot\boldsymbol e_f)]_i$. With $\chi=(\phi-\delta)/s$, the normal force on a face is

\begin{equation}\label{eq:csf}
f_{\sigma,d,f}=\sigma\overline\kappa_f\frac{\phi_R-\phi_L}{s h},
\qquad
f_{\sigma,d,i}=\frac12\left(f_{\sigma,d,i+\frac12\boldsymbol e_d}
+f_{\sigma,d,i-\frac12\boldsymbol e_d}\right).
\end{equation}

The cell momentum increment is $\Delta t\boldsymbol f_{\sigma,i}$. Averaging the face forces before dividing by density keeps the force placement consistent with the cell momentum update. Surface tension is applied in a separate explicit stage, and its finite kinetic-energy change is included in the budget. The resulting pressure--capillary balance and parasitic currents are assessed in \ref{app:static_drop}.

\subsection{Transport predictor and compatible explicit pressure relaxation}\label{sec:pressure}

Time advancement is fully explicit. Each physical time step consists of an explicit transport predictor, a separate capillary update when surface tension is present, and a fixed number of local pressure-relaxation iterations. The pressure gradient at the beginning of the step is included in the predictor, and the final pressure increment corrects momentum once. This sequence couples pressure to centered transport without an implicit pressure or velocity solve.

Let $q=(\phi,m_1,m_2,\boldsymbol M)$ and let $L(q;U^n,p^n)$ contain the phase equations and the momentum right-hand side $\boldsymbol R_M-G_cp^n$, with $\boldsymbol R_M$ defined in Eq.~\eqref{eq:momentum_flux}. The pressure at the beginning of the step therefore acts throughout the momentum predictor. Its physical-cell values and the advecting face velocity $U^n$ are held fixed during the three-stage strong-stability-preserving (SSP) Runge--Kutta (RK) update \cite{Gottlieb2001}:

\begin{equation}\label{eq:ssprk}
\begin{aligned}
q^{[1]}&=q^n+\Delta t L(q^n;U^n,p^n),\\
q^{[2]}&=\frac34q^n+\frac14\{q^{[1]}+\Delta t L(q^{[1]};U^n,p^n)\},\\
q^{\mathrm{tr}}&=\frac13q^n+\frac23\{q^{[2]}+\Delta t L(q^{[2]};U^n,p^n)\}.
\end{aligned}
\end{equation}

Phase fluxes, intrinsic densities, cell velocities in the momentum flux, and stresses are evaluated at each stage. We denote the predictor state by $q^{\mathrm{tr}}$ and set $\rho^*=m_1^{\mathrm{tr}}+m_2^{\mathrm{tr}}$. Including the existing pressure force in these stages couples the pressure-supported motion to the transport. Pressure relaxation then supplies the increment needed to update that force. The pressure iterations are performed once after the predictor.

Define $A$ as arithmetic interpolation of cell velocity to face-normal velocity, $D_c=DA$, and $G_c$ as the centered cell pressure gradient. For a uniform periodic mesh, $G_c=-D_c^\dagger$ under the cell-volume inner product $\langle a,b\rangle_V=\sum_i V_i a_i\cdot b_i$. The pressure is first evolved using the incoming face velocity:

\begin{equation}\label{eq:p_initial}
p^{(0)}=p^n-\rho^*c^2\Delta t\,DU^n.
\end{equation}

When surface tension is present, the capillary stage gives $\boldsymbol M^*=\boldsymbol M^{\mathrm{tr}}+\Delta t\boldsymbol f_\sigma$; otherwise $\boldsymbol M^*=\boldsymbol M^{\mathrm{tr}}$. During pressure relaxation, $\rho^*$, $\phi^{\mathrm{tr}}$, and $\boldsymbol M^*$ remain fixed. With the pressure increment $\pi=p-p^n$, the trial velocity is

\begin{equation}\label{eq:trial_velocity}
\pi=p-p^n,\qquad
\boldsymbol u(p)=\frac{\boldsymbol M^*-\Delta tG_c\pi}{\rho^*},
\qquad U(p)=A\boldsymbol u(p).
\end{equation}

\begin{equation}\label{eq:p_relax}
p^{(r+1)}=p^{(r)}-\rho^*c^2\Delta t\,D U(p^{(r)}),
\qquad r=0,\ldots,N_p-1.
\end{equation}

We use $N_p=20$ additional iterations following Eq.~\eqref{eq:p_initial}. Each iteration updates the pressure boundary values and evaluates the trial velocity from the same predicted momentum and the current pressure increment. The resulting increment is applied to momentum once:

\begin{equation}\label{eq:final_update}
\begin{aligned}
p^{n+1}&=p^{(N_p)},\qquad
\boldsymbol M^{n+1}=\boldsymbol M^*-\Delta tG_c(p^{n+1}-p^n),\\
\boldsymbol u^{n+1}&=\frac{\boldsymbol M^{n+1}}{\rho^*},
\qquad U^{n+1}=A\boldsymbol u^{n+1}.
\end{aligned}
\end{equation}

The face flux used by the next transport step therefore corresponds to the corrected cell momentum. Each relaxation update evaluates its right-hand side from the preceding pressure trial; the iteration index does not represent an additional physical time step. The fixed iteration count introduces no implicit solve or elliptic convergence tolerance. Distributed execution exchanges neighboring pressure data at each iteration. The kinetic-energy interpretation and step-size condition for this operator pairing are developed in Section~\ref{sec:pressure_property}.

\subsection{Time step, boundary conditions, and single-phase limit}\label{sec:time}

The explicit updates are subject to acoustic, interface-regularization, viscous, and capillary time-step restrictions. For grid spacing $h$, the time step is selected as

\begin{equation}\label{eq:dt}
\Delta t\le\min\left\{
\frac{C_a h}{U_{\max}+c},\quad
\frac{C_d h^2}{\Gamma\epsilon},\quad
\frac{C_d h^2}{\nu_{\max}},\quad
C_\sigma\sqrt{\frac{\min_\ell\rho_\ell^0 h^3}{\pi\sigma}}
\right\}.
\end{equation}

Terms with absent diffusion or surface tension are omitted. Here $U_{\max}$ is the largest absolute cell or face velocity component and $\nu_{\max}$ is the maximum evaluated diffusivity, including SGS viscosity in three dimensions. The two-dimensional benchmarks of Sections~\ref{sec:single} and~\ref{sec:two} use $C_d=C_\sigma=0.1$ and $C_a=0.1$, except for cavities with $C_a=0.3$. The HIT calculations use $(C_a,C_d)=(0.2,0.08)$, and three-dimensional dam breaks use $(C_a,C_d,C_\sigma)=(0.1,0.08,0.1)$. Prescribed time-step ceilings are applied in addition to these bounds.

The role of the Runge--Kutta stages can be seen from constant-coefficient advection on a periodic grid. Centered differencing gives imaginary eigenvalues $\mathrm{i}\omega$, and the stability polynomial of Eq.~\eqref{eq:ssprk}, $R(z)=1+z+z^2/2+z^3/6$, satisfies $|R(\mathrm{i}\omega\Delta t)|\le1$ for $|\omega\Delta t|\le\sqrt{3}$. Thus centered transport can be advanced explicitly without adding an upwind dissipative flux. This linear restriction complements the diffusion and capillary bounds in Eq.~\eqref{eq:dt}; the variable-density pressure iterations have the separate condition in Eq.~\eqref{eq:spectral_condition}, and phase-mass positivity is addressed in Section~\ref{sec:h_properties}. Centered and WENO3-JS comparisons use the same time-step policy (Section~\ref{sec:comparison}).

The reference artificial Mach number is $0.02$ in the two-dimensional benchmarks of Sections~\ref{sec:single} and~\ref{sec:two}, and $0.03$ in the three-dimensional dam break. Shear layers and cavities use $c=50$. For bubbles and the two-dimensional dam break, $c=\max(U_{\mathrm{ref}}/0.02,U_{\max}/0.02)$. The three-dimensional dam break uses $c=\max(U_{\mathrm{ref}}/0.03,1.1U_*/0.03)$, where $U_*$ is the larger of $U_{\max}$ and the largest cell velocity magnitude. The regularization speed is $\Gamma=\gamma_\Gamma U_{\max}$, with the problem-dependent coefficient given in Table~\ref{tab:numerics}. These acoustic speeds control the numerical pressure evolution; they are not physical sound speeds for acoustic-wave prediction.

Periodic boundaries share fluxes across opposing faces. At a solid wall, normal phase-volume and phase-mass fluxes vanish, and the specified no-slip or slip condition is imposed on velocity. The phase fraction has zero normal gradient, corresponding to neutral wetting. Pressure ghost values satisfy the hydrostatic normal gradient $\partial_n p=\rho\boldsymbol g\cdot\boldsymbol n$, which reduces to a homogeneous Neumann condition in the absence of normal gravity. During the predictor, the ghost values use the current stage density while the interior values of $p^n$ remain fixed. Momentum and velocity boundary values are imposed consistently before each stage.

Single-phase calculations use $\phi=1$, $m_1=\rho$, and $m_2=0$. The phase-fraction equation and interface-regularization terms are omitted, and the mass flux is $F_{m,f}=\overline\rho_f U_f$. Thus the single-phase scheme uses the arithmetic mean of the fluid density, whereas the two-phase scheme uses Eq.~\eqref{eq:harmonic} for each intrinsic phase density. Both use the same centered momentum flux and EPP update. The two-phase formulation is common to two and three spatial dimensions.

%% file: properties.tex
\section{Discrete transport and pressure properties}\label{sec:properties}

The analysis below establishes the properties of the three main components of the method. Centered momentum transport cancels both numerical production and dissipation of kinetic energy by advection. Harmonic density interpolation bounds phase-mass outflow and satisfies an inverse-density transport identity without altering that cancellation. Compatible pressure operators yield a conditional energy estimate for successive relaxation iterates. Section~\ref{sec:budget} then combines the operator work and finite-stage energy changes into the budget used in the numerical tests.

\subsection{Conservative transport and advective kinetic-energy cancellation}\label{sec:ke_identity}

Sharing each face flux makes the cell sum of either phase-mass right-hand side telescope. The phase integrals are consequently conserved for periodic or impermeable boundaries, up to floating-point accumulation. The same argument gives conservative momentum advection. The additional kinetic-energy property follows from the velocity used in the momentum flux.

\begin{equation}\label{eq:ke_chain}
K=\sum_i V_i\frac{|\boldsymbol M_i|^2}{2\rho_i},\qquad
\dot K=\sum_i V_i\left(\boldsymbol u_i\cdot\dot{\boldsymbol M}_i
-\frac12|\boldsymbol u_i|^2\dot\rho_i\right).
\end{equation}

For an internal face oriented from $L$ to $R$, its advective contribution is

\begin{equation}\label{eq:ke_face}
\dot K_f=S_fF_{m,f}\left[
(\boldsymbol u_R-\boldsymbol u_L)\cdot\overline{\boldsymbol u}_f
-\frac{|\boldsymbol u_R|^2-|\boldsymbol u_L|^2}{2}\right]=0.
\end{equation}

This facewise identity uses the common mass flux and arithmetic velocity average. It eliminates both numerical production and numerical dissipation of kinetic energy by spatial advection, without requiring a divergence-free velocity or a particular density mean. In particular, harmonic phase-density interpolation retains the identity with the same ACDI mass transport that appears in the phase equations. Molecular and nonnegative eddy-viscosity stresses retain their dissipative role, as shown in \ref{app:viscous}. Pressure, surface tension, gravity, and finite time increments contribute separately to the coupled energy balance. Their contributions are measured in Section~\ref{sec:budget}, allowing the effect of upwind momentum reconstruction to be identified under otherwise common conditions.

\subsection{Harmonic phase density: outflow bound and inverse-density identity}\label{sec:h_properties}

The ratio $r_\ell=m_\ell/\alpha_\ell$ can develop a large local contrast when a phase occupies only a small fraction of a cell. An arithmetic intrinsic-density mean, $r^A_{\ell,f}=(r_{\ell,L}+r_{\ell,R})/2$, then transfers a neighboring large density into the outgoing mass flux of a low-density cell. The harmonic mean instead satisfies $0<r^H_{\ell,f}\le2r_{\ell,i}$ for either adjacent cell with positive intrinsic densities. This is a symmetric property of the mean, independent of the direction of transport.

To relate this face bound to one explicit update, define the fraction of the cell's phase volume carried outward during $\Delta t$ by

\begin{equation}\label{eq:phase_cfl}
C_{\alpha_\ell,i}=\frac{\Delta t}{V_i\alpha_{\ell,i}}
\sum_{f\in\partial i}S_f\max(s_{if}F_{\alpha_\ell,f},0).
\end{equation}

For a forward-Euler (FE) mass update with positive input $m_\ell$ and $\alpha_\ell$, incoming contributions are nonnegative. Bounding the outgoing harmonic density by $2r_{\ell,i}$ gives

\begin{equation}\label{eq:positivity}
\begin{aligned}
m_{\ell,i}^{\mathrm{FE}}
&\ge m_{\ell,i}-\frac{\Delta t}{V_i}
\sum_f S_f\max(s_{if}F_{\alpha_\ell,f},0)\,2r_{\ell,i}\\
&=m_{\ell,i}(1-2C_{\alpha_\ell,i}).
\end{aligned}
\end{equation}

Hence $C_{\alpha_\ell,i}<1/2$ is a sufficient condition for positive phase mass in a forward-Euler update. Since Eq.~\eqref{eq:ssprk} is a convex combination of these updates, positivity is retained when this condition and positive phase fractions hold at every constituent stage. Here $C_{\alpha_\ell,i}$ uses the actual phase-volume flux, including ACDI regularization. Finite velocity divergence affects that flux and the evolving phase fraction, but introduces no source in the conservative phase-mass update. Section~\ref{sec:3d_positivity} compares this sufficient condition with the measured stage minima in the dam break.

A complementary semi-discrete identity describes why the harmonic mean is suited to independently evolved phase volumes and masses. Consider an inverse-density functional that grows when phase mass approaches zero while phase volume remains finite:

\begin{equation}\label{eq:barrier}
\mathcal B_\ell=\sum_i V_i\frac{\alpha_{\ell,i}^2}{m_{\ell,i}}
=\sum_i V_i\frac{\alpha_{\ell,i}}{r_{\ell,i}}.
\end{equation}

The harmonic mean satisfies $r^H_{\ell,f}(1/r_{\ell,L}+1/r_{\ell,R})=2$. Substituting this relation into the internal-face contribution obtained by differentiating Eq.~\eqref{eq:barrier} gives

\begin{equation}\label{eq:barrier_face}
S_fF_{\alpha_\ell,f}\left(\frac1{r_{\ell,R}}-\frac1{r_{\ell,L}}\right)
\left[2-r^H_{\ell,f}\left(\frac1{r_{\ell,L}}+\frac1{r_{\ell,R}}\right)\right]=0.
\end{equation}

This cancellation follows specifically from harmonic intrinsic-density interpolation. With the arithmetic mean, the square bracket in Eq.~\eqref{eq:barrier_face} becomes $-(r_{\ell,R}-r_{\ell,L})^2/(2r_{\ell,L}r_{\ell,R})$, which is nonzero for unequal adjacent densities. Thus, whereas the advective kinetic-energy identity is independent of the density mean, the inverse-density identity uses the harmonic mean essentially. Summing over cells under periodic or no-flux boundary conditions therefore yields

\begin{equation}\label{eq:barrier_identity}
\frac{d\mathcal B_\ell}{dt}
=2\sum_i V_i\frac{\alpha_{\ell,i}}{r_{\ell,i}}\theta_i.
\end{equation}

Harmonic interpolation cancels the internal-face contribution of the full phase-volume flux, including interface regularization, leaving only the finite-divergence source. No assumption of $\theta=0$ is needed. Consequently, even when the intrinsic density varies across the interface, transport with the harmonic mean produces no change in $\mathcal B_\ell$. Its evolution is governed by the resolved compression and expansion. This identity describes the spatial transport structure; Eq.~\eqref{eq:positivity} provides the corresponding control of mass loss over a finite explicit stage.

The mean remains a centered, second-order face approximation for a smooth positive intrinsic density. In fact,

\begin{equation}\label{eq:mean_difference}
r^H_f-r^A_f=-\frac{(r_R-r_L)^2}{4r^A_f}.
\end{equation}

Consequently, the two means have the same linearization about a constant intrinsic density, and differ by $O(h^2)$ for smooth data. The robustness mechanism is therefore compatible with the centered transport design. Section~\ref{sec:3d_positivity} tests this mechanism in the three-dimensional dam break using matched computations.

\subsection{Kinetic-energy behavior of pressure relaxation}\label{sec:pressure_property}

We next examine the kinetic-energy response of the local pressure iterations. The pressure update uses the cell gradient that corrects momentum and the divergence of its interpolated velocity. Fix the predicted momentum $\boldsymbol M^*$ and positive density $\rho^*$ on a uniform periodic grid, as in the relaxation stage of Section~\ref{sec:pressure}. The kinetic energy of the trial velocity is then a quadratic function of the pressure increment $\pi=p-p^n$:

\begin{equation}\label{eq:pressure_functional}
\mathcal K(\pi)=\frac12\left\langle
\boldsymbol M^*-\Delta tG_c\pi,
R^{-1}(\boldsymbol M^*-\Delta tG_c\pi)\right\rangle_V,
\qquad R=\operatorname{diag}(\rho^*).
\end{equation}

The adjoint relation gives $\nabla_\pi\mathcal K=\Delta tD_c\boldsymbol u(p)$. With $B=\operatorname{diag}(\rho^*c^2)$, each additional iteration is therefore the preconditioned descent step

\begin{equation}\label{eq:descent}
\pi^{(r+1)}=\pi^{(r)}-B\nabla_\pi\mathcal K(\pi^{(r)}).
\end{equation}

\begin{equation}\label{eq:spectral_condition}
\lambda_{\max}\!\left(\Delta t^2 B^{1/2}G_c^\dagger R^{-1}G_cB^{1/2}\right)<2
\quad\Longrightarrow\quad
\mathcal K(\pi^{(r+1)})\le\mathcal K(\pi^{(r)}).
\end{equation}

The derivation in \ref{app:pressure} identifies how the acoustic step and the density distribution enter this condition. Under the condition, each additional iteration reduces or leaves unchanged the kinetic energy of the preceding trial velocity. This establishes an energy-control property of the compatible pressure operators, including nonuniform positive density. The comparison is between successive pressure trials at one predicted state. The change between physical time levels also includes predictor pressure work and the initial pressure update in Eq.~\eqref{eq:p_initial}; it is evaluated through the complete budget in Section~\ref{sec:budget}.

%% file: verification.tex
\section{Verification design and fully discrete energy budget}\label{sec:verification}

\subsection{Matched reconstruction comparisons and reference problems}\label{sec:comparison}

The assessment combines controlled energy tests, incompressible reference problems, and large-deformation benchmarks. Inviscid homogeneous isotropic turbulence (HIT) tests centered advection in single- and two-phase flow and isolates the pressure--transport coupling. Periodic shear layers relate energy loss to vortical transport, while cavities compare wall-bounded velocity profiles with incompressible reference data. Rising bubbles test capillary and buoyant deformation. Two- and three-dimensional dam breaks then examine impact, entrainment, phase-mass robustness, and the energy pathways under an explicit SGS model.

To quantify the benefit of eliminating upwind momentum dissipation, we construct a comparison scheme by replacing only the centered velocity $\overline{\boldsymbol u}_f$ in Eq.~\eqref{eq:momentum_flux} with third-order weighted essentially non-oscillatory (WENO) reconstruction using Jiang--Shu (JS) nonlinear weights \cite{JiangShu1996}. This comparison scheme is denoted WENO3-JS; the proposed method retains centered reconstruction throughout. The sign of $F_{m,f}$ selects the WENO stencil, specified in \ref{app:weno}. Phase transport, EPP, molecular stress, surface tension, time-step policy, and SGS model are common to the two calculations. Differences in the measured advective work can therefore be attributed to momentum reconstruction. Arithmetic/harmonic comparisons in Sections~\ref{sec:hit_mean} and \ref{sec:3d_positivity} evaluate the choice of intrinsic-density mean in the phase-mass flux.

\begin{table}[htbp]
\centering\small
\caption{Physical parameters. Paired values refer to phases 1 and 2; single-phase rows list the sole fluid value. Viscosities are dynamic, and dimensional two-phase benchmarks use SI units.}\label{tab:physical}
\begin{tabular}{@{}>{\raggedright\arraybackslash}p{.25\linewidth}rrrr@{}}\toprule
Problem & $\rho_1^0,\rho_2^0$ & $\mu_1,\mu_2$ & $g_y$ & $\sigma$\\\midrule
Single-phase HIT & 1 & 0 & 0 & 0\\
Two-phase HIT & 1000, 1 & 0, 0 & 0 & 0\\
Single-phase shear layer & 1 & $0$ or $10^{-4}$ & 0 & 0\\
Cavity & 1 & $1/Re$ & 0 & 0\\
Two-phase shear layer & 100, 1 & $10^{-2},10^{-4}$ & 0 & 0\\
Bubble A & 1000, 100 & 10, 1 & $-0.98$ & 24.5\\
Bubble B & 1000, 1 & 10, 0.1 & $-0.98$ & 1.96\\
2D dam break & 1000, 1 & $10^{-3},1.8\times10^{-5}$ & $-9.81$ & 0\\
3D dam break & 1000, 1 & $10^{-3},1.8\times10^{-5}$ & $-9.8$ & 0.0728\\
\bottomrule
\end{tabular}
\end{table}

\begin{table}[htbp]
\centering\small
\caption{Interface and time-integration parameters. $n$ denotes cells per direction in HIT and per unit length in shear-layer tests. Listed $\Delta t$ values are time-step ceilings. All cases use 20 additional pressure iterations.}\label{tab:numerics}
\begin{tabular}{@{}>{\raggedright\arraybackslash}p{.36\linewidth}rrl@{}}\toprule
Problem & $\epsilon/h$ & $\gamma_\Gamma$ & $\Delta t$\\\midrule
HIT, $n=64,128$ & 0.51 (two-phase) & 1.0192 & $2\times10^{-5}(64/n)$\\
Single-phase shear layer & -- & -- & $5\times10^{-6}(128/n)$\\
Cavity, $Re=1000,5000$ & -- & -- & Adaptive\\
Two-phase shear layer & 0.51 & 1.0192 & $5\times10^{-6}$\\
Bubbles A/B & 0.6792465 & 4.368 & Adaptive\\
2D dam break & 0.6792465 & 1.0192 & Adaptive\\
3D dam break & 0.51 & 1.0192 & Adaptive\\
\bottomrule
\end{tabular}
\end{table}

Kinetic energy is evaluated using the actual mixture density, with $K_0=K(0)$ denoting its initial value. In periodic two-dimensional tests, enstrophy is $Z=\tfrac12\sum_iV_i\omega_{z,i}^2$, with $\omega_z=\partial_xv-\partial_yu$ evaluated by centered differences. Velocity divergence is evaluated as

\begin{equation}\label{eq:divdiagnostic}
(\nabla\cdot\boldsymbol u)_i=
\frac1{V_i}\sum_{f\in\partial i}s_{if}S_fU_f.
\end{equation}

Integral diagnostics use cell-volume weights. Reference errors are evaluated at the published sampling locations, or by linear interpolation of simulation histories to reference times. The relative discrete $L_2$ difference is

\begin{equation}\label{eq:l2}
e_2(q)=\left[\frac{\sum_k(q_k-q_k^{\mathrm{ref}})^2}
{\sum_k(q_k^{\mathrm{ref}})^2}\right]^{1/2}.
\end{equation}

\subsection{Operator work and finite-stage energy changes}\label{sec:budget}

We evaluate kinetic energy from the transported mass and momentum using Eq.~\eqref{eq:ke_chain}. At every predictor stage, the contribution of each right-hand-side term is measured before the Runge--Kutta combination. The advective power is

\begin{equation}\label{eq:power}
P_{\mathrm{adv}}=\sum_iV_i\left[
\boldsymbol u_i\cdot(-D\boldsymbol F_M)_i
-\frac12|\boldsymbol u_i|^2(-DF_m)_i\right].
\end{equation}

For periodic boundaries, the same quantity has the face representation

\begin{equation}\label{eq:adv_power}
P_{\mathrm{adv}}=\sum_fS_fF_{m,f}(\boldsymbol u_R-\boldsymbol u_L)\cdot
\left(\boldsymbol u_f^M-\overline{\boldsymbol u}_f\right).
\end{equation}

The momentum-transport velocity $\boldsymbol u_f^M$ is the arithmetic mean in Centered and the reconstructed velocity in WENO3-JS. Interface-regularization transport is included through the common mass flux $F_m$. For first-order upwind momentum transport, Eq.~\eqref{eq:adv_power} gives a nonpositive face contribution proportional to the squared velocity jump, whereas the centered contribution vanishes. For the nonlinear WENO3-JS reconstruction, the signed work is evaluated directly from Eq.~\eqref{eq:adv_power}.

Molecular viscosity, SGS stress, and gravity are evaluated from their respective momentum right-hand sides. The predictor pressure power is

\begin{equation}\label{eq:work_weights}
\begin{gathered}
P_{p,\mathrm{RK}}^{[s]}=\sum_iV_i\boldsymbol u_i^{[s]}\cdot(-G_cp^n)_i,
\\ W_j^n=\Delta t\sum_{s=0}^2b_s P_j^{[s]},
\quad (b_0,b_1,b_2)=\left(\frac16,\frac16,\frac23\right).
\end{gathered}
\end{equation}

Here $q^{[0]}=q^n$, and $j$ denotes advection, molecular viscosity, SGS stress, gravity, or predictor pressure. The pressure force is the one actually applied at each stage, including the boundary extension. On a uniform periodic grid, the adjoint relation gives $P_{p,\mathrm{RK}}^{[s]}=\langle p^n,D_c\boldsymbol u^{[s]}\rangle_V$: its sign is set by the correlation between pressure and velocity divergence. The transport time residual is the difference between the finite predictor energy change and its accumulated right-hand-side work:

\begin{equation}\label{eq:temporal_residual}
R_{\mathrm{tr,time}}^n=K(q^{\mathrm{tr}})-K(q^n)
-W_{\mathrm{adv}}^n-W_\mu^n-W_{\mathrm{SGS}}^n-W_g^n-W_{p,\mathrm{RK}}^n.
\end{equation}

At the final pressure correction, density is fixed and $\delta\boldsymbol M_p=-\Delta tG_c(p^{n+1}-p^n)$. Its exact finite kinetic-energy change is

\begin{equation}\label{eq:pressure_energy}
\Delta K_{p,\mathrm{corr}}^n=\sum_iV_i\left[
\frac{\boldsymbol M_i^*\cdot\delta\boldsymbol M_{p,i}}{\rho_i^*}
+\frac{|\delta\boldsymbol M_{p,i}|^2}{2\rho_i^*}\right].
\end{equation}

In terms of Section~\ref{sec:pressure_property}, this correction is $\Delta K_{p,\mathrm{corr}}^n=\mathcal K(\pi^{(N_p)})-\mathcal K(0)$, whereas Eq.~\eqref{eq:spectral_condition} compares consecutive iterates starting from $\pi^{(0)}=p^{(0)}-p^n$. The complete pressure contribution over the physical step is the sum of the predictor work and the final correction:

\begin{equation}\label{eq:pressure_total}
W_{p,\mathrm{total}}^n=W_{p,\mathrm{RK}}^n+\Delta K_{p,\mathrm{corr}}^n.
\end{equation}

The capillary contribution $\Delta K_\sigma^n$ is likewise the measured energy difference across the CSF stage. Summing over time steps gives

\begin{equation}\label{eq:full_budget}
\begin{aligned}
K(t)-K(0)={}&W_{\mathrm{adv}}+W_\mu+W_{\mathrm{SGS}}+W_g\\
&+W_{p,\mathrm{RK}}+\Delta K_{p,\mathrm{corr}}+\Delta K_\sigma
+R_{\mathrm{tr,time}}+\varepsilon_K.
\end{aligned}
\end{equation}

The closure residual $\varepsilon_K$ verifies the accounting of the recorded updates. The term $R_{\mathrm{tr,time}}$ measures the predictor's discrete energy-chain residual, with all force work already removed. Pressure work can have either sign; its total and its two components are reported with their measured signs. Two-dimensional energies are per unit depth. The three-dimensional dam break starts at rest, so its energies are reported in joules.

OpenAI ChatGPT and Codex assisted with the analysis and visualization of numerical results using the recorded simulation data and the diagnostic definitions given above.

%% file: hit.tex
\section{Inviscid homogeneous isotropic turbulence}\label{sec:hit}

\subsection{Configurations and comparison design}\label{sec:hit_setup}

Inviscid HIT tests the energy retained by the coupled transport and pressure algorithm in the absence of molecular viscosity and SGS dissipation. We compare centered and WENO3-JS momentum transport in single-phase flow and in a density-ratio-$1000$ two-phase flow. The two-phase configuration follows the spherical-drop test of Hwang and Jain \cite{HwangJain2024}. A separate comparison changes only the intrinsic-density mean in the centered two-phase method.

The domain is a periodic cube of side $2\pi L_0$, with $L_0=1$. The initial velocity has zero spatial mean and the target spectrum

\begin{equation}\label{eq:hit_scales}
\begin{gathered}
E(k)\propto k^4\exp\!\left[-2\left(\frac{k}{k_0}\right)^2\right],
\qquad k_0L_0=4,\\
U_0=\sqrt{\frac{\langle|\boldsymbol u(0)|^2\rangle}{3}}=1,
\qquad t^*=\frac{tU_0}{L_0}.
\end{gathered}
\end{equation}

Fourier velocity modes are projected perpendicular to the discrete wave vector $\widetilde k_d=\sin(k_dh)/h$ and normalized to $U_0=1$; zero and Nyquist-plane modes are omitted. A fixed pseudorandom realization (seed 20261118) generates one initial velocity file per resolution, shared by all comparisons on that grid. Initially, $p=1$ and the single-phase density is $\rho=1$. The two-phase case adds a spherical liquid drop of radius $L_0$ at the domain center, with $\rho_1^0/\rho_2^0=1000/1$ and the finite-bound interface profile of Section~\ref{sec:acdi}. The initial velocity is retained inside and outside the drop. Gravity and surface tension are zero.

Centered and WENO3-JS are compared on $64^3$ and $128^3$ cells through $t^*=8$; harmonic averaging is used in the two-phase cases. The energy budgets and interface visualization below use $128^3$ cells, with an additional centered arithmetic-mean case on the same grid. All cases start with $c=640$ and use 20 additional pressure iterations. The sound-speed controller uses a local Mach-number limit of $0.1$ and safety factor $1.1$; $c$ remains 640 in these calculations. Time-step ceilings and interface parameters are listed in Table~\ref{tab:numerics}. Within each comparison, the initial field, pressure treatment, time-step policy, and interface settings are common.

\subsection{Kinetic-energy retention in single-phase and two-phase flow}\label{sec:hit_energy}

\paperfig{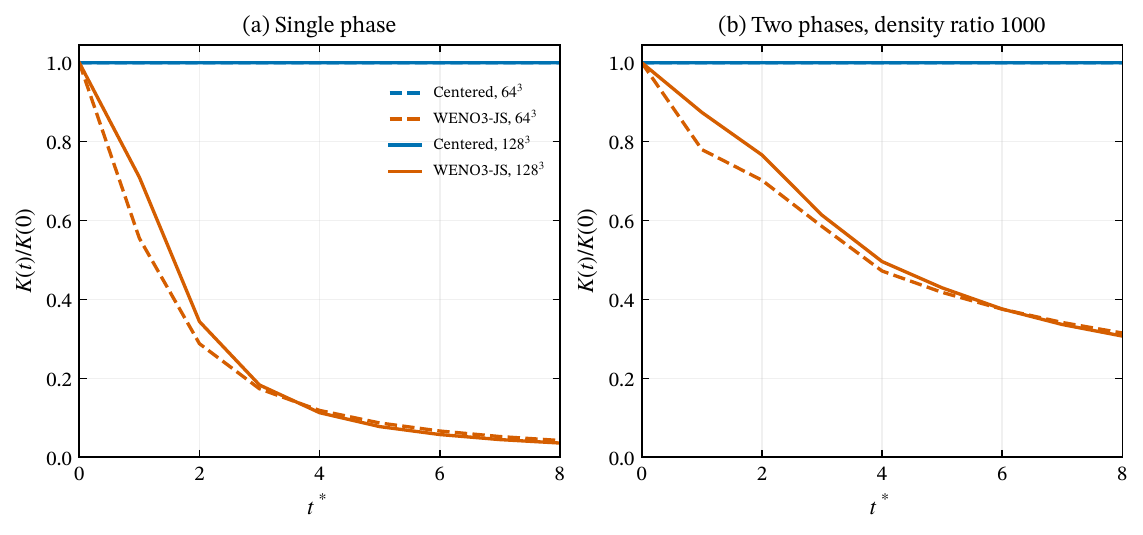}{Normalized kinetic energy in inviscid HIT. (a) Single phase; (b) two phases at density ratio 1000 with harmonic intrinsic-density averaging. Solid curves show $128^3$ cells and dashed curves $64^3$ cells. Blue and orange denote centered and WENO3-JS momentum transport. Each case is normalized by its own $K_0$; both panels share the same vertical scale.}{fig:hit_refinement}

Figure~\ref{fig:hit_refinement} shows a clear separation between centered energy retention and upwind decay in both single-phase and two-phase flow. On $128^3$ cells at $t^*=8$, the relative energy changes of the centered method are $-7.47\times10^{-8}$ and $-1.18\times10^{-7}$, respectively, whereas WENO3-JS retains $3.60\%$ and $30.72\%$ of the initial energies. The same separation is present on $64^3$ cells: centered energy changes remain within $2.2\times10^{-7}$ in magnitude, while WENO3-JS retains $4.24\%$ and $31.49\%$. Removing upwind momentum dissipation therefore preserves the initial kinetic energy to a high degree in both configurations and on both grids.

\subsection{Advective and pressure contributions to the energy budget}\label{sec:hit_budget}

Table~\ref{tab:hit_energy} resolves the $128^3$ energy changes into cumulative advective work, total pressure work, and the transport time residual. The centered advective contribution is negligible at the scale of the measured energy change, whereas WENO3-JS loses $96.40\%$ and $69.28\%$ of $K_0$ through advection in single-phase and two-phase flow. Thus momentum reconstruction accounts for the large difference in Fig.~\ref{fig:hit_refinement} under the common pressure treatment.

\begin{table}[htbp]
\centering\small
\caption{Inviscid HIT energy changes and cumulative work at $t^*=8$ on $128^3$ cells, normalized by each case's $K_0$. SP and TP denote single-phase and two-phase flow; C and W denote centered and WENO3-JS transport; H and AR denote harmonic and arithmetic intrinsic-density means. The last row provides the density-mean comparison in Section~\ref{sec:hit_mean}.}\label{tab:hit_energy}
\begin{tabular}{@{}>{\raggedright\arraybackslash}p{.14\linewidth}rrrr@{}}\toprule
Case & $\Delta K/K_0$ & $W_{\mathrm{adv}}/K_0$ & $W_{p,\mathrm{total}}/K_0$ & $R_{\mathrm{tr,time}}/K_0$\\\midrule
SP C & $-7.472\times10^{-8}$ & $2.780\times10^{-20}$ & $-7.461\times10^{-8}$ & $-1.037\times10^{-10}$\\
SP W & $-9.640\times10^{-1}$ & $-9.640\times10^{-1}$ & $-3.941\times10^{-8}$ & $-6.799\times10^{-13}$\\
TP C H & $-1.185\times10^{-7}$ & $-4.268\times10^{-18}$ & $-1.045\times10^{-7}$ & $-1.396\times10^{-8}$\\
TP W H & $-6.928\times10^{-1}$ & $-6.928\times10^{-1}$ & $-2.643\times10^{-8}$ & $7.636\times10^{-12}$\\
TP C AR & $-1.201\times10^{-7}$ & $6.643\times10^{-19}$ & $-1.060\times10^{-7}$ & $-1.408\times10^{-8}$\\
\bottomrule
\end{tabular}
\end{table}

For the centered cases, Fig.~\ref{fig:hit_pressure} shows predictor pressure work and the final increment correction separately. In two-phase flow, the predictor contributes $+3.36\times10^{-7}K_0$ and the final correction $-4.40\times10^{-7}K_0$. Their sum, together with the transport time residual, accounts for the small net energy loss. The net pressure work on $64^3$ cells is $+2.27\times10^{-7}K_0$. This positive contribution is likewise small compared with the upwind advective loss. The estimate in Section~\ref{sec:pressure_property} applies to successive pressure iterations at a fixed predictor; Eqs.~\eqref{eq:pressure_energy}--\eqref{eq:full_budget} account for the complete physical step, including the predictor pressure work, whose sign can vary.

\paperfig{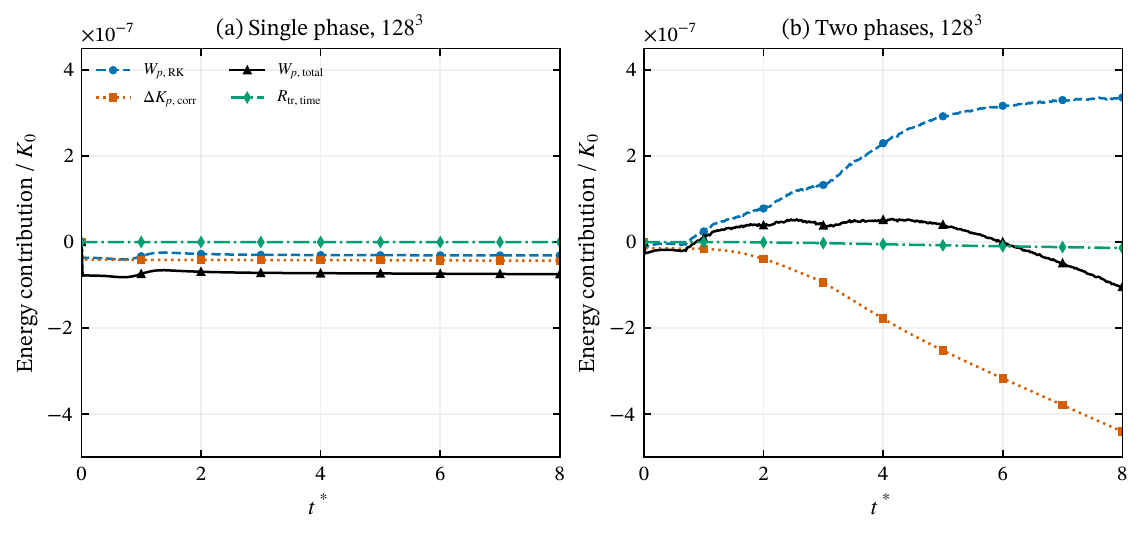}{Energy-budget contributions in centered HIT on $128^3$ cells. Predictor pressure work, final pressure-increment correction, their sum, and the transport time residual are normalized by $K_0$. (a) Single phase; (b) two phases with harmonic averaging. Both panels use the same vertical scale; lines connect sampled values, with markers at integer times.}{fig:hit_pressure}

For the centered $128^3$ single-phase and two-phase harmonic calculations, the maximum normalized closure residuals are $3.81\times10^{-16}$ and $5.69\times10^{-16}$; the corresponding WENO3-JS values are $4.05\times10^{-12}$ and $1.19\times10^{-12}$. Maximum local Mach numbers in the centered cases remain below $0.014$ and $0.090$, respectively. A separate pressure-coupling comparison in \ref{app:coupling} holds the spatial operators and acoustic settings fixed. Including the existing pressure in the predictor and applying only the final increment reduces the energy-change magnitude by more than four orders relative to applying the full pressure after a pressure-free predictor. The pressure connection therefore helps retain the benefit of centered transport in the complete calculation.

\subsection{Intrinsic-density averaging in the two-phase calculation}\label{sec:hit_mean}

To isolate the density mean, we compare the centered $128^3$ harmonic and arithmetic cases with identical initial data, interface parameters, and pressure settings. Both reach $t^*=8$ with positive phase masses. Their relative kinetic-energy changes are $-1.18\times10^{-7}$ and $-1.20\times10^{-7}$, respectively, and both retain the advective cancellation in Table~\ref{tab:hit_energy}. These results support the compatibility of harmonic phase-mass transport with kinetic-energy-preserving momentum advection. The stronger phase-mass robustness test is the large-deformation dam break in Section~\ref{sec:3d_positivity}, where the density means produce different outcomes.

Figure~\ref{fig:hit_interface} shows the interface evolution in the same centered harmonic calculation on $128^3$ cells. The initial sphere develops elongated and fragmented structures by $t^*=8$. The small net kinetic-energy change is thus maintained through substantial interface deformation.

\paperfig{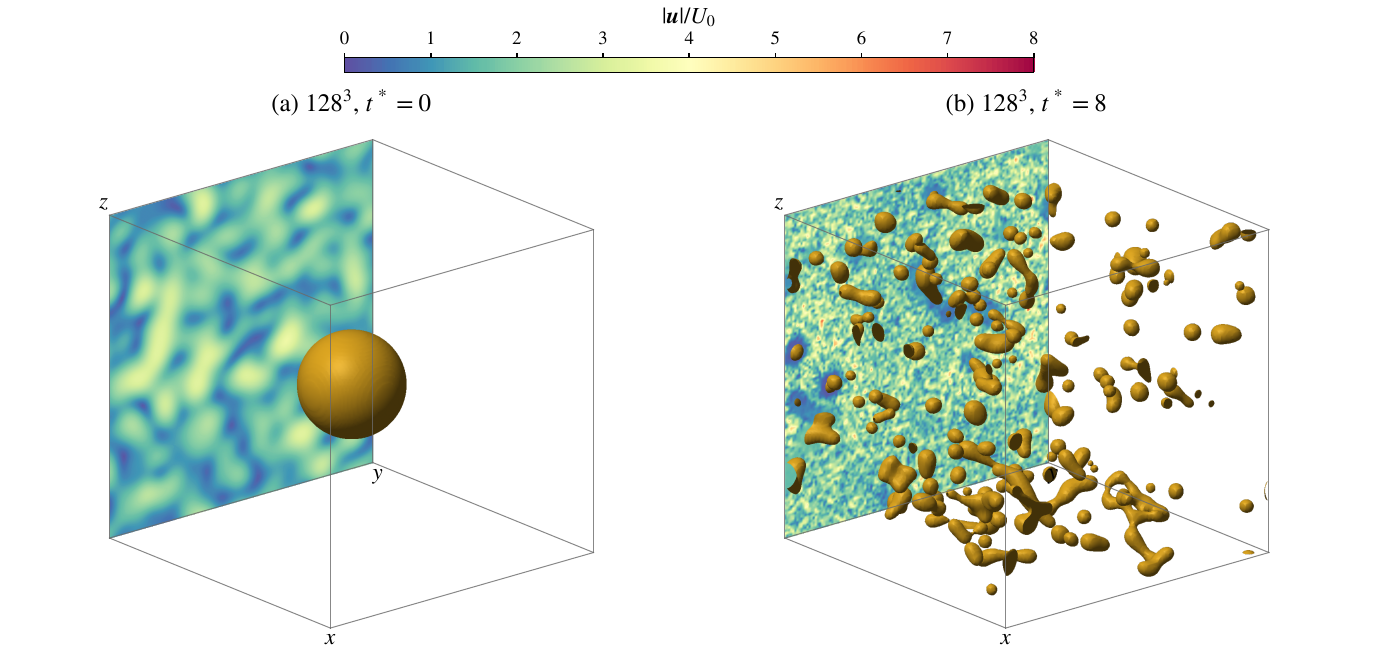}{Interface evolution in centered two-phase inviscid HIT on $128^3$ cells with harmonic intrinsic-density averaging. (a) $t^*=0$; (b) $t^*=8$. The gold surface denotes $\phi=0.5$; the colored $x=0$ plane shows $|\boldsymbol u|/U_0$ after interpolating cell velocity vectors to grid points. Both panels use the same viewpoint and color scale.}{fig:hit_interface}

%% file: single_phase.tex
\section{Single-phase verification}\label{sec:single}

\subsection{Doubly periodic shear layer}\label{sec:single_dpsl}

The doubly periodic shear layer provides a controlled test of vortical transport without wall work. The domain is $[0,1]^2$, with initial density $\rho=1$ and reference scales $U=L=1$. The initial velocity is

\begin{equation}\label{eq:dpsl_initial}
u(x,y)=\begin{cases}
\tanh[80(y-0.25)],&y<0.5,\\
\tanh[80(0.75-y)],&y\ge0.5,
\end{cases}\qquad
v(x,y)=0.05\sin[2\pi(x+0.25)].
\end{equation}

Calculations use $128^2$ and $256^2$ cells to $tU/L=1$, with $c=50$ and $\Delta t=5\times10^{-6}(128/n)$. Both inviscid flow and $\nu=10^{-4}$ are considered. The pressure treatment is identical for Centered and WENO3-JS, so that the inviscid comparison isolates the effect of momentum reconstruction on the energy budget.

\paperfig{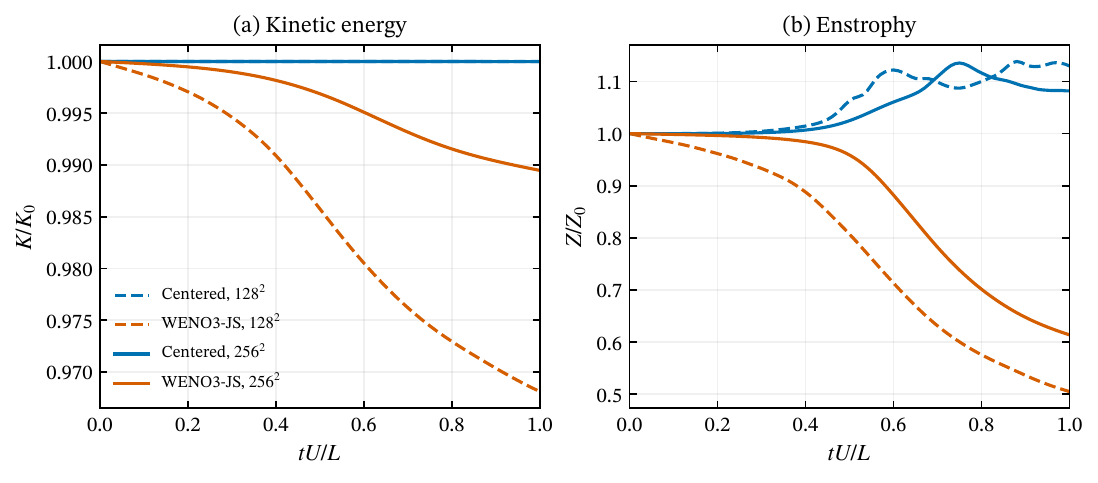}{Inviscid single-phase shear layer: normalized kinetic energy and enstrophy. Color identifies momentum reconstruction and line style identifies resolution.}{fig:single_inviscid}

Figure~\ref{fig:single_inviscid} and Table~\ref{tab:single_inviscid} show a clear separation between the two transport operators. On $128^2$ cells, the centered energy loss is $0.000889\%$ of $K_0$, compared with $3.18914\%$ for WENO3-JS. The centered advective work is $7.12\times10^{-17}\%$ of $K_0$, whereas WENO3-JS removes $3.18838\%$ through advection. The total pressure contributions are $-0.000889\%$ and $-0.000754\%$, respectively. The large difference in energy retention therefore originates in momentum advection under common EPP conditions.

\begin{table}[htbp]
\centering\small
\caption{Single-phase inviscid energy budget at $tU/L=1$, in percent of $K_0$; $n$ denotes $n^2$ cells.}\label{tab:single_inviscid}
\begin{tabular}{@{}lrrrr@{}}\toprule
Method, $n$ & $\Delta K$ & $W_{\mathrm{adv}}$ & $W_\mu$ & $W_{p,\mathrm{total}}$\\\midrule
Centered, 128 & -0.000889 & $7.119\times10^{-17}$ & 0.000000 & -0.000889\\
WENO3-JS, 128 & -3.189138 & $-3.188\times10^{0}$ & 0.000000 & -0.000754\\
Centered, 256 & -0.000966 & $5.526\times10^{-17}$ & 0.000000 & -0.000966\\
WENO3-JS, 256 & -1.052334 & $-1.051\times10^{0}$ & 0.000000 & -0.000885\\
\bottomrule
\end{tabular}
\end{table}

On $256^2$ cells, the centered energy loss remains below $0.001\%$ of $K_0$, whereas the WENO3-JS loss is $1.05233\%$. Refinement reduces the upwind loss, but its advective contribution remains more than three orders of magnitude larger than the total pressure contribution. Both grids therefore show the benefit of eliminating the advective energy sink.

\paperfig{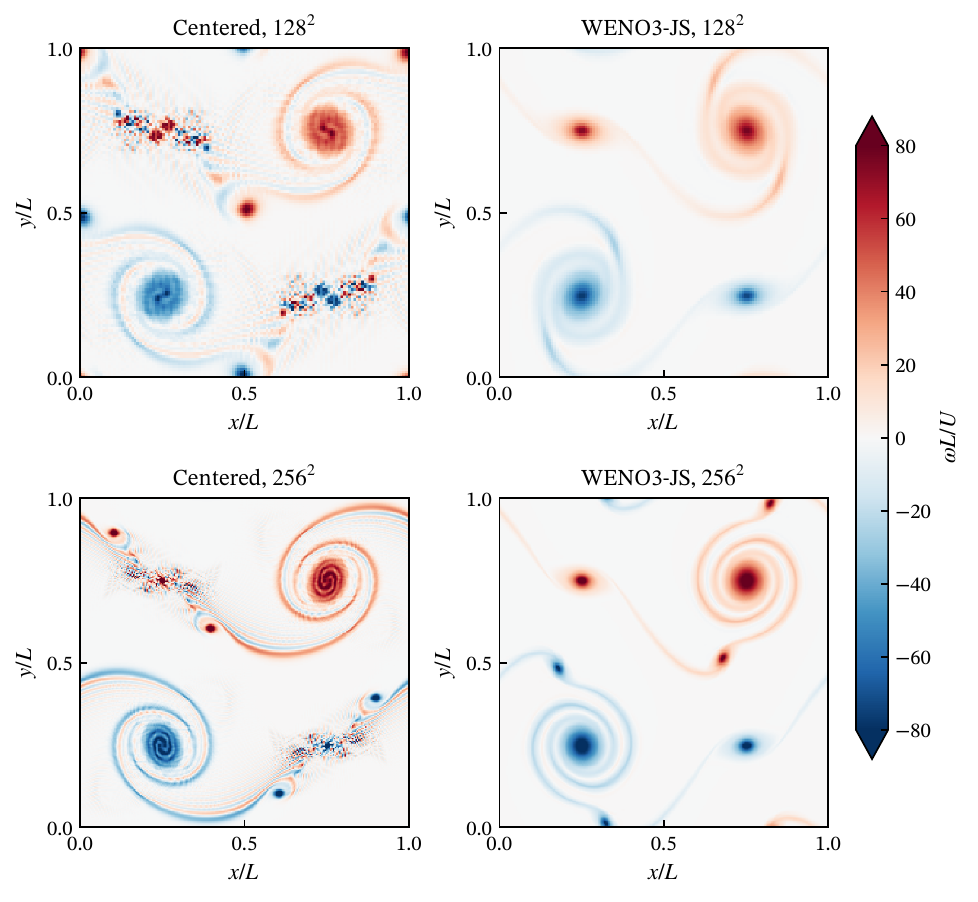}{Single-phase inviscid vorticity at $tU/L=1$. Rows: $128^2$ and $256^2$; columns: Centered and WENO3-JS. All panels show the full unit domain with one color range.}{fig:omega_inviscid}

Figure~\ref{fig:omega_inviscid} shows how the difference in energy loss affects the vortical structures. The centered method retains tightly wound shear layers and stronger vorticity variations that are damped by the upwind reconstruction. Enstrophy distinguishes the methods more strongly than total kinetic energy because it weights velocity gradients. This inviscid test demonstrates the transport property and its effect on the retained structures; the finite-viscosity case provides the complementary test with a prescribed physical dissipation scale.

\paperfig{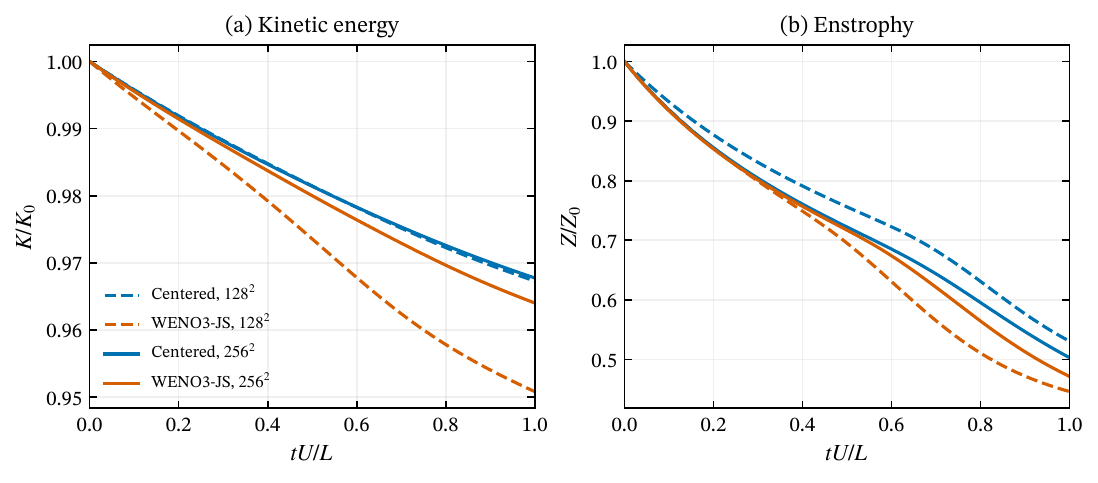}{Single-phase shear layer at $\nu=10^{-4}$: normalized kinetic energy and enstrophy, with the same reconstruction and grid conventions as Fig.~\ref{fig:single_inviscid}.}{fig:single_viscous}

\begin{table}[htbp]
\centering\small
\caption{Single-phase viscous energy budget at $tU/L=1$ and $\nu=10^{-4}$, in percent of $K_0$.}\label{tab:single_viscous}
\begin{tabular}{@{}lrrrr@{}}\toprule
Method, $n$ & $\Delta K$ & $W_{\mathrm{adv}}$ & $W_\mu$ & $W_{p,\mathrm{total}}$\\\midrule
Centered, 128 & -3.269966 & $2.552\times10^{-11}$ & -3.269147 & -0.000819\\
WENO3-JS, 128 & -4.917211 & $-1.980\times10^{0}$ & -2.936504 & -0.000696\\
Centered, 256 & -3.224370 & $-1.372\times10^{-10}$ & -3.223528 & -0.000842\\
WENO3-JS, 256 & -3.597175 & $-4.318\times10^{-1}$ & -3.164519 & -0.000811\\
\bottomrule
\end{tabular}
\end{table}

Figure~\ref{fig:single_viscous} and Table~\ref{tab:single_viscous} show that centered transport retains the prescribed viscous decay at $\nu=10^{-4}$. On $128^2$ cells, molecular viscosity accounts for $3.26915\%$ of $K_0$, compared with a total pressure contribution of $-0.000819\%$. WENO3-JS removes a further $1.98001\%$ through momentum advection, while its viscous loss is $2.93650\%$. The latter changes because the common stress operator acts on different resolved gradients. Eliminating upwind damping therefore allows the prescribed viscosity to determine dissipation through the retained flow field.

\paperfig{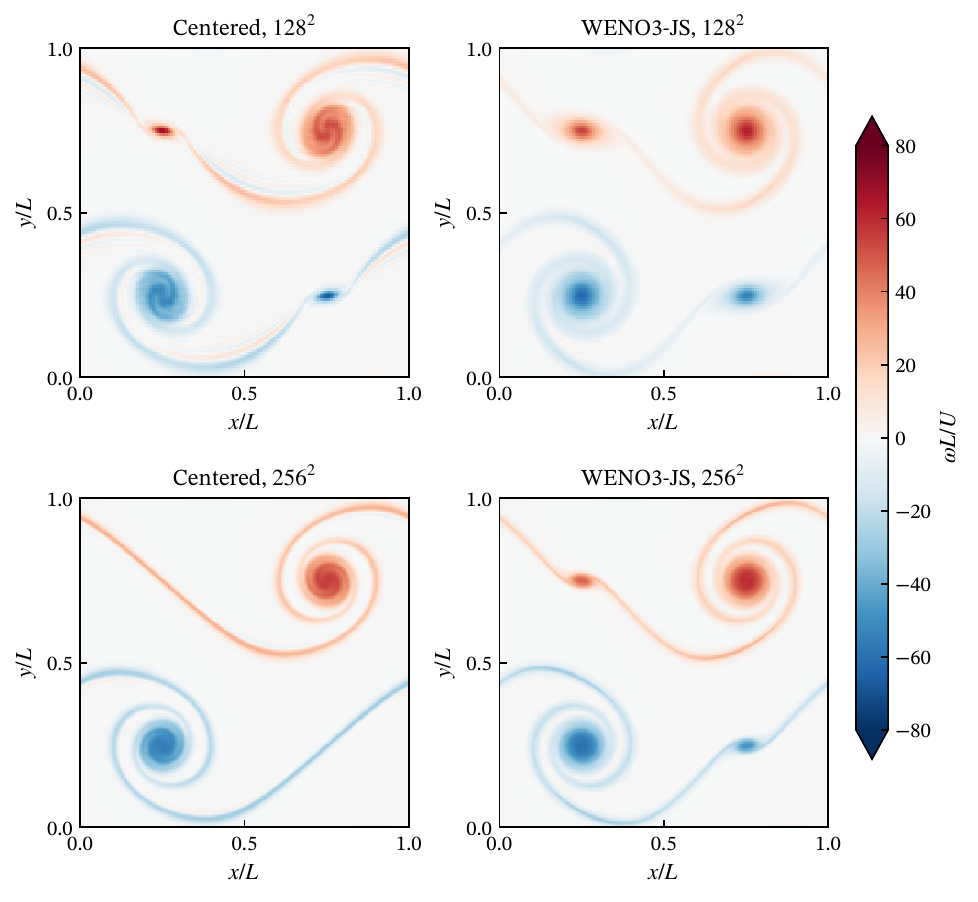}{Single-phase vorticity at $tU/L=1$ for $\nu=10^{-4}$, on the full unit domain. The same color range is used in all four panels.}{fig:omega_viscous}

The vorticity comparison in Fig.~\ref{fig:omega_viscous} links the energy budget to the resolved shear layers. At the same molecular viscosity, centered advection retains sharper gradients than the upwind reconstruction. On $256^2$ cells, centered viscous loss is $3.22353\%$ of $K_0$ and total pressure work is $-0.000842\%$. The WENO3-JS advective loss falls to $0.431845\%$, yet still amounts to $13.6\%$ of its molecular-viscous loss. The difference in transport dissipation therefore persists under refinement.

\subsection{Lid-driven cavity}\label{sec:cavity}

The unit-square cavity has a lid velocity $U=1$ and stationary no-slip conditions on the other walls. With $\rho=1$ and $\mu=1/Re$, the cases $Re=1000$ and $5000$ are computed on $128^2$ and $256^2$ grids. Figure~\ref{fig:cavity} compares the centered and WENO3-JS velocity profiles with the 17 sampling points per component reported by Ghia et al.\ \cite{Ghia1982}.

\paperfig{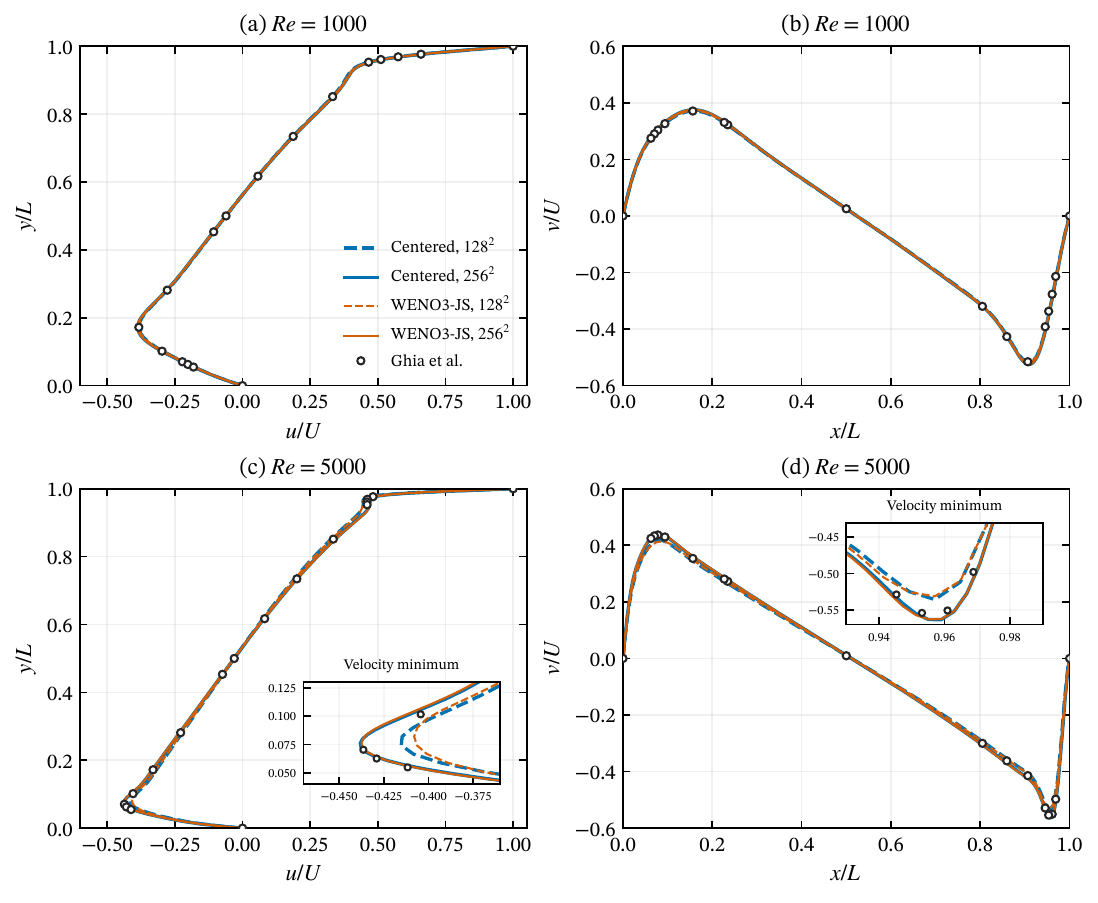}{Lid-driven cavity centerline profiles: $u(y)$ at $x/L=0.5$ and $v(x)$ at $y/L=0.5$. Top: $Re=1000$; bottom: $Re=5000$. Blue and orange curves denote centered and WENO3-JS transport; dashed and solid curves show $128^2$ and $256^2$ cells, respectively. Insets enlarge the velocity minima at $Re=5000$. Circles are the incompressible reference data of Ghia et al.\ \cite{Ghia1982}.}{fig:cavity}

The centered profiles reproduce the principal recirculation and near-wall velocity extrema. At $Re=5000$, refinement reduces the maximum absolute differences at the reference samples from $0.0307$ to $0.00874$ for $u/U$ and from $0.0271$ to $0.00899$ for $v/U$. Thus the centered weakly compressible method captures the established incompressible circulation without upwind stabilization of momentum transport.

Velocity variation is monitored between outputs one convective time apart, using $\max|\Delta u_a|L/(U^2\Delta t_{\mathrm{out}})$ and a threshold of $10^{-6}$. All eight centered and WENO3-JS calculations satisfy this criterion. Under the same grid, pressure, and viscosity settings, the WENO3-JS profiles also approach the reference solution; their maximum absolute differences at $Re=5000$ on $256^2$ cells are $0.0109$ for $u/U$ and $0.0102$ for $v/U$.

%% file: two_phase.tex
\section{Two-phase verification}\label{sec:two}

\subsection{Viscous two-phase shear layer}\label{sec:two_dpsl}

The periodic velocity field of Eq.~\eqref{eq:dpsl_initial} is combined with two liquid disks of radius $0.15$, centered at $(0.25,0.25)$ and $(0.75,0.75)$. The reference density ratio is $100$, with $(\mu_1,\mu_2)=(0.01,10^{-4})$, so both phases have reference kinematic viscosity $10^{-4}$. The Reynolds number based on $U=L=1$ is $10^4$. Gravity and surface tension are absent. The phase fraction is initialized from the periodic signed distance, and phase masses use their reference intrinsic densities. Calculations on $128^2$ and $256^2$ cells use $\Delta t=5\times10^{-6}$ and the harmonic phase-density mean.

\paperfig{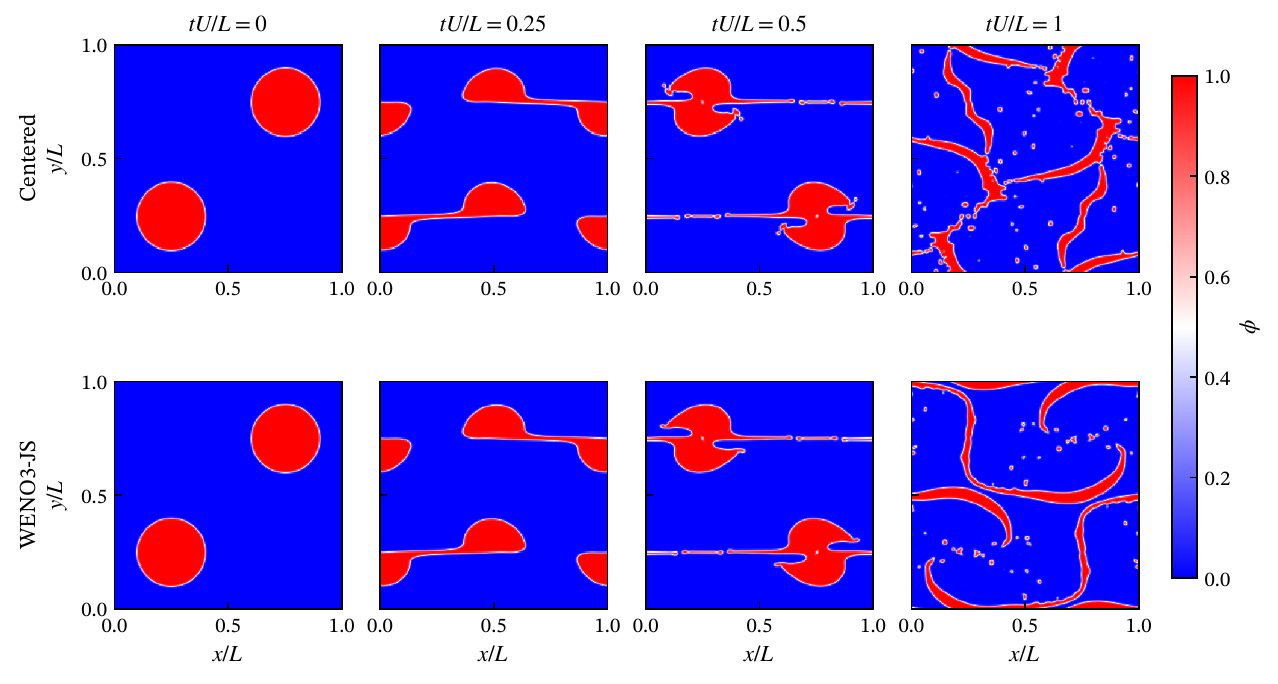}{Two-phase shear-layer phase fraction on $256^2$ cells. Top: Centered; bottom: WENO3-JS. The displayed times are $tU/L=0$, $0.25$, $0.5$, and $1$. The color scale spans $\phi=0$ (gas) to $\phi=1$ (liquid).}{fig:two_phi}

\paperfig{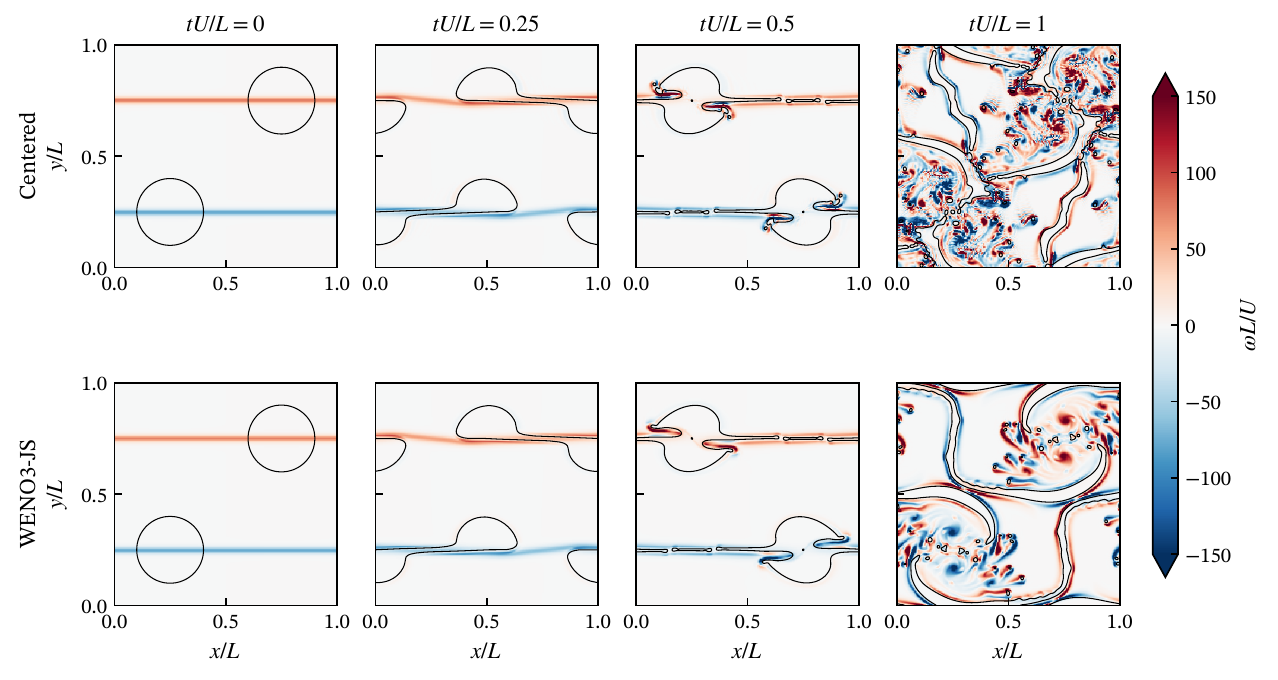}{Vorticity and the $\phi=0.5$ interface for the same two-phase shear layer. One vorticity scale is used for both reconstructions and all times.}{fig:two_omega}

The early disk stretching is similar for the two momentum reconstructions (Fig.~\ref{fig:two_phi}). As the layers elongate and interact, Centered retains finer vortical motion around the deformed interfaces, whereas WENO3-JS gives a smoother velocity field (Fig.~\ref{fig:two_omega}). Because the density and viscosity vary with the phase distribution, this difference also changes the molecular-viscous work. The energy decomposition is consequently more informative than the final kinetic energy alone.

\paperfig{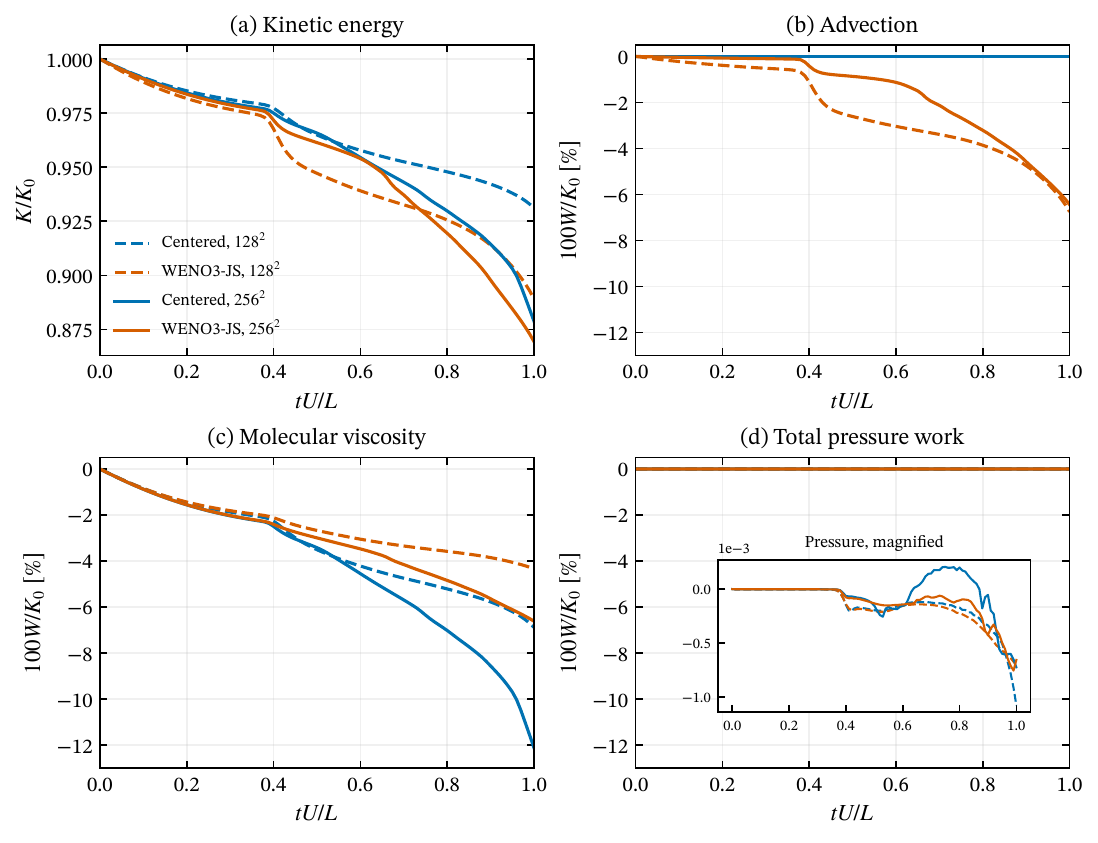}{Two-phase shear-layer energy histories. Panels (b)--(d) use the same vertical range for cumulative advection, molecular viscosity, and pressure work, all expressed as a percentage of $K_0$.}{fig:two_budget}

\begin{table}[htbp]
\centering\small
\caption{Two-phase shear-layer cumulative work at $tU/L=1$, in percent of $K_0$. The maximum closure error normalized by $K_0$ is below $8.8\times10^{-13}$.}\label{tab:two_budget}
\begin{tabular}{@{}lrrrr@{}}\toprule
Method, $n$ & $W_{\mathrm{adv}}$ & $W_\mu$ & $W_{p,\mathrm{total}}$ & $R_{\mathrm{tr,time}}$\\\midrule
Centered, 128 & $9.531\times10^{-12}$ & -6.879596 & -0.001079 & $1.436\times10^{-10}$\\
WENO3-JS, 128 & $-6.748\times10^{0}$ & -4.334879 & -0.000708 & $1.076\times10^{-10}$\\
Centered, 256 & $-5.687\times10^{-9}$ & -12.110751 & -0.000726 & $-5.197\times10^{-9}$\\
WENO3-JS, 256 & $-6.440\times10^{0}$ & -6.614254 & -0.000658 & $2.125\times10^{-9}$\\
\bottomrule
\end{tabular}
\end{table}

Table~\ref{tab:two_budget} confirms that harmonic intrinsic-density averaging retains the centered transport property in variable-density flow. At $256^2$, the measured centered advective work is $-5.69\times10^{-9}\%$ of $K_0$, compared with $-6.44015\%$ for WENO3-JS. Molecular viscosity removes $12.11075\%$ and $6.61425\%$, respectively. Centered transport thus allows the stronger resolved gradients to feed the explicitly prescribed viscous dissipation, while the upwind reconstruction removes a substantial fraction directly through momentum advection.

The total centered pressure contribution decreases in magnitude from $0.001079\%$ to $0.000726\%$ of $K_0$ on refinement. On the finer grid, it is $0.0060\%$ of the molecular-viscous loss. This quantity includes the pressure work in the transport stages as well as the final increment correction. The common scale in Fig.~\ref{fig:two_budget} shows that pressure relaxation preserves the practical low-dissipation advantage: molecular viscosity remains the principal energy sink for the centered calculation.

All four harmonic shear-layer calculations retain positive phase masses at the constituent Euler stages. The positivity condition is evaluated using the actual phase-volume flux, including ACDI regularization, as in Eq.~\eqref{eq:positivity}. This links the measured mass transport to the same flux that appears in the kinetic-energy identity.

\subsection{Rising bubbles}\label{sec:bubbles}

The rising-bubble benchmarks of Hysing et al.\ \cite{Hysing2009} use a $1\times2$ domain, with a gas bubble of radius $0.25$ initially centered at $(0.5,0.5)$. The fluid is initially at rest. The vertical boundaries are free-slip and the horizontal boundaries are no-slip. Gravity is $g_y=-0.98$. Case A has density ratio $10$ and surface tension $24.5$; case B has density ratio $1000$ and surface tension $1.96$. The physical parameters are given in Table~\ref{tab:physical}. Case A uses $64\times128$ and $128\times256$ cells, and case B uses $128\times256$ and $256\times512$ cells. Both are run to $t=3$ s.

\begin{equation}\label{eq:bubble_metrics}
y_c=\frac{\sum_iV_i(1-\phi_i)y_i}{\sum_iV_i(1-\phi_i)},\qquad
V_b=\frac{\sum_iV_i(1-\phi_i)v_i}{\sum_iV_i(1-\phi_i)}.
\end{equation}

The centroid $y_c$ and gas-volume-weighted rise velocity $V_b$ are compared with the finest published TP2D, FreeLIFE, and MooNMD histories. Final interfaces use the $\phi=0.5$ contour. The reference solutions differ most in case B, where the interface undergoes stronger deformation.

\paperfig{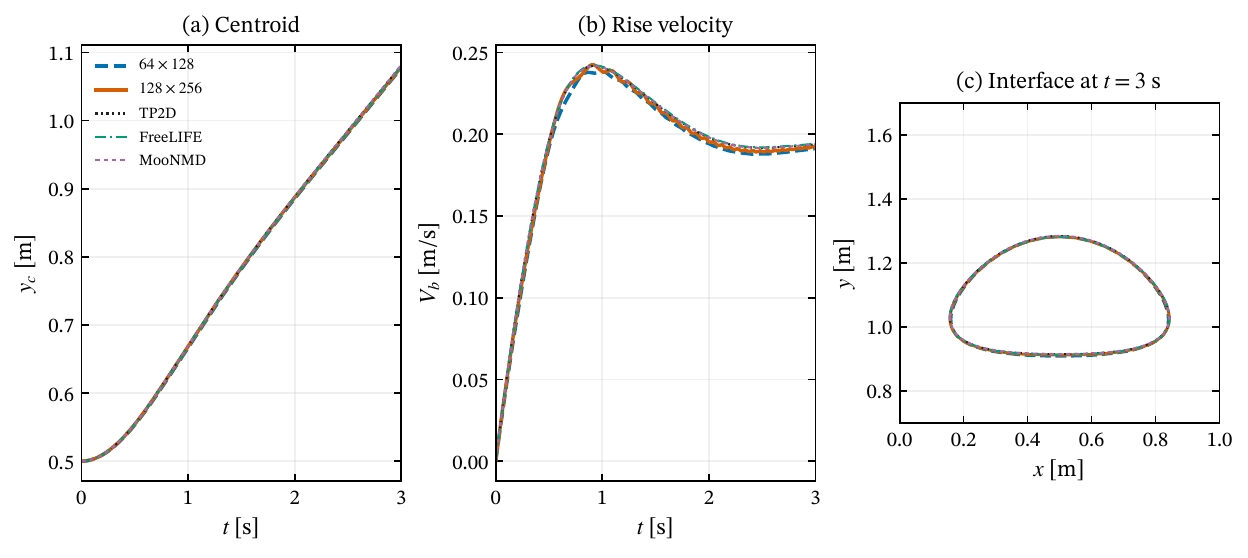}{Rising-bubble case A: centroid, rise velocity, and final interface. References are the finest TP2D, FreeLIFE, and MooNMD data of Hysing et al.\ \cite{Hysing2009}.}{fig:bubble_A}

Case A shows close agreement in both the rise history and the deformed bubble cap (Fig.~\ref{fig:bubble_A}). On the $128\times256$ grid, the relative $L_2$ differences from TP2D over $t=0.01$--$3$ s are $0.149\%$ for centroid height and $1.03\%$ for rise velocity. The finite-width interface retains the principal bubble shape while reproducing the acceleration and subsequent leveling of its rise speed.

\paperfig{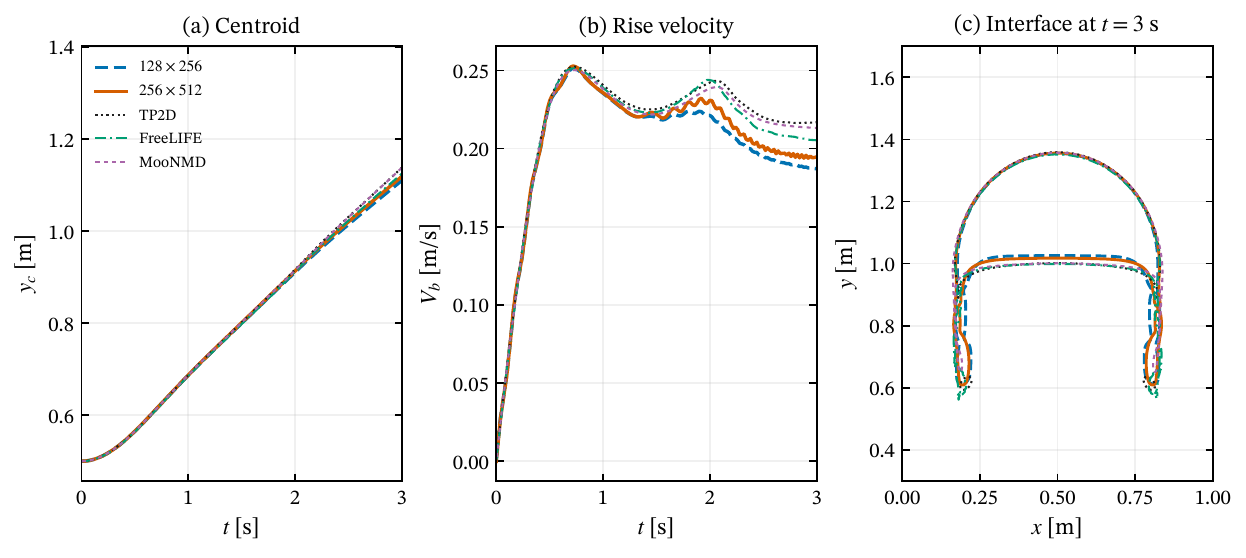}{Rising-bubble case B, with the same diagnostics as Fig.~\ref{fig:bubble_A}. The final interfaces show the bubble cap and the thin trailing structures.}{fig:bubble_B}

Case B develops a pronounced skirt and thin trailing structures (Fig.~\ref{fig:bubble_B}). The computation captures the main upward displacement and cap shape, with the trajectory approaching the TP2D trend as the grid is refined. The relative $L_2$ differences from TP2D decrease from $1.227\%$ to $0.806\%$ for centroid height and from $7.44\%$ to $5.52\%$ for rise velocity. The later evolution follows the TP2D tendency within the spread associated with different treatments of the strongly deformed interface.

The two cases exercise the same centered momentum and harmonic phase-mass fluxes without a change of transport treatment at the larger density ratio. The comparisons demonstrate that the centered formulation captures buoyant rise and capillary deformation at both density ratios.

\subsection{Two-dimensional dam break}\label{sec:dam2d}

A rectangular water column of width $a=0.146$ m and height $H_0=0.292$ m collapses in a $0.584\times0.584$ m tank. All walls, including the roof, are no-slip. The initial velocity is zero, $g=9.81$ m/s$^2$, and surface tension is omitted. The reference velocity is $U_{\mathrm{ref}}=\sqrt{gH_0}=1.69248$ m/s. Calculations on $128^2$ and $256^2$ cells extend to $t=1$ s, covering the initial surge, downstream-wall impact, run-up, and subsequent return flow.

\paperfig{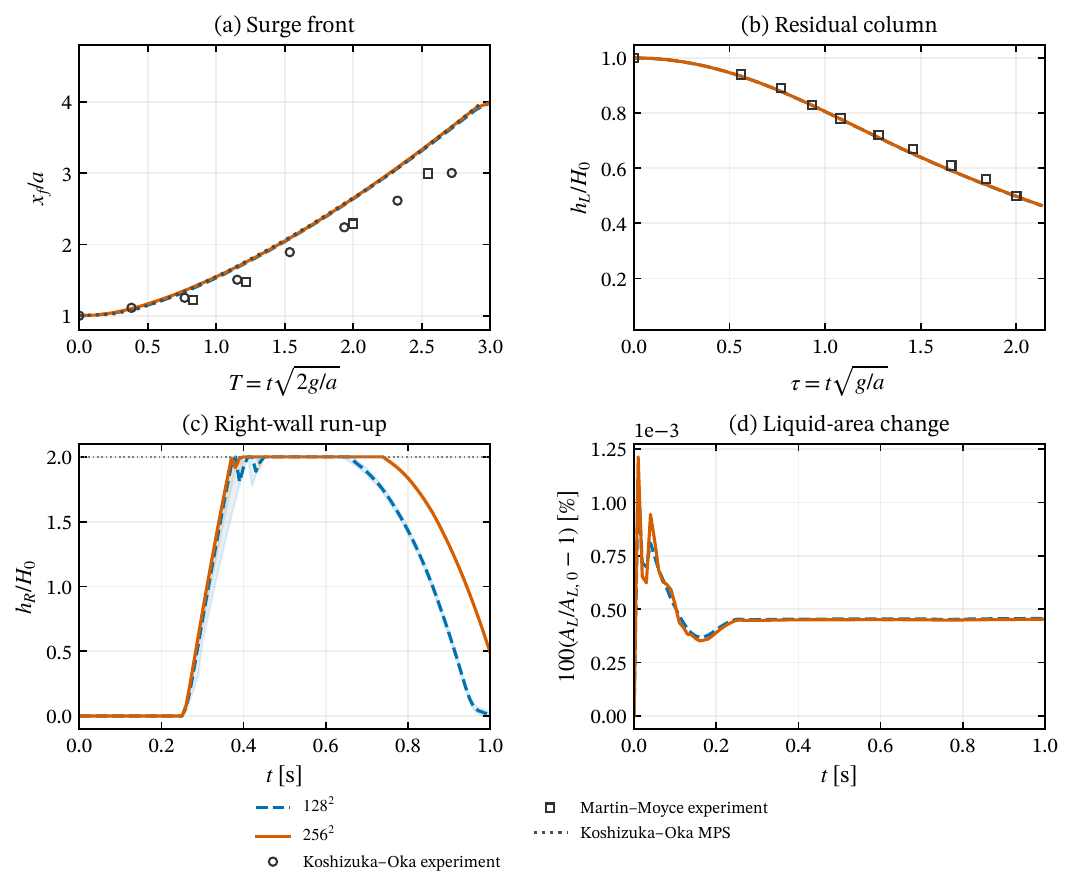}{Two-dimensional dam-break histories. (a) Surge front compared with the Martin--Moyce and Koshizuka--Oka experimental points and the Koshizuka--Oka moving-particle semi-implicit (MPS) computation. (b) Residual-column height compared with Martin--Moyce. (c) Highest wet point on the right wall; shading spans the $\phi=0.1$ and $0.9$ crossings. (d) Relative liquid-area change. Experimental comparisons concern the pre-impact interval.}{fig:dam2d_history}

Figure~\ref{fig:dam2d_history} compares normalized surge-front positions with the experiments of Martin and Moyce \cite{MartinMoyce1952} and Koshizuka and Oka \cite{KoshizukaOka1996}, together with the published moving-particle semi-implicit (MPS) computation. The front follows the MPS propagation, while the residual-column height closely follows the Martin--Moyce measurements. The experimental front positions tend to lag the numerical curves in the later pre-impact interval. These comparisons cover both horizontal spreading and vertical collapse of the initial column.

\paperfig{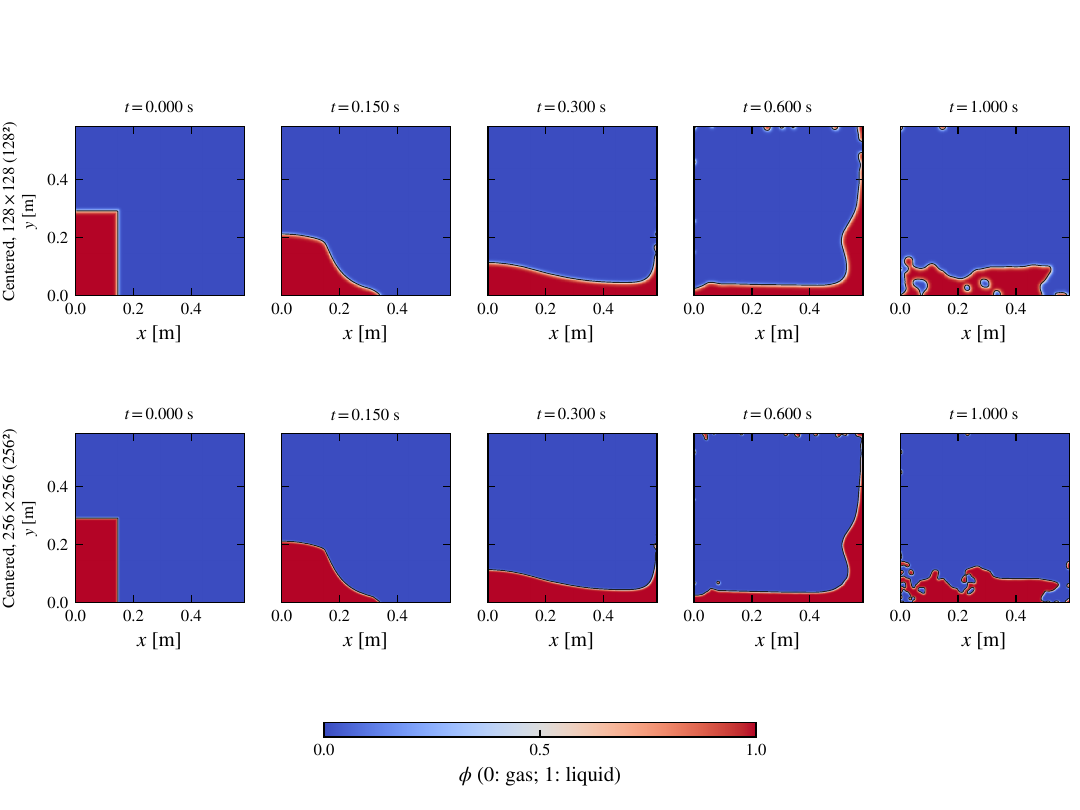}{Two-dimensional dam-break phase evolution on $128^2$ and $256^2$ cells. The black line is $\phi=0.5$; the actual saved times are shown. The domain includes the closed roof. Top: $128^2$; bottom: $256^2$. The common color scale denotes the liquid volume fraction.}{fig:dam2d_shapes}

After impact, the liquid runs up the downstream wall, reaches the roof, and folds back into the tank (Fig.~\ref{fig:dam2d_shapes}). The right-wall history records this run-up and return, while the liquid-area history quantifies the small volume response of the weakly compressible formulation. The calculation follows this strongly deforming flow with the same centered transport used in the periodic and buoyancy-driven tests.

Both grids retain positive phase masses through $t=1$ s. The maximum relative mass errors are $5.80\times10^{-13}$ and $1.24\times10^{-12}$ on $128^2$ and $256^2$ cells, respectively. Figure~\ref{fig:dam2d_history} compares the front, wall run-up, and liquid-area histories at both resolutions.

%% file: three_dimensional.tex
\section{Three-dimensional dam break with an explicit SGS model}\label{sec:three}

\subsection{Configuration and subgrid-scale closure}\label{sec:3dsetup}

A wet-floor dam break provides a three-dimensional application involving rapid interface deformation, wall impact, and entrainment. Such configurations have also been used in weakly compressible and lattice Boltzmann multiphase studies \cite{Matsushita2021,SitompulAoki2019}. The present tank dimensions are $L_x\times L_y\times L_z=0.72\times0.36\times0.12$ m. A water column of width $0.15$ m occupies the full height $0.36$ m at the left end, and the rest of the floor is initially covered by a $0.018$ m liquid layer. All walls, including the roof and spanwise walls, are no-slip. The initial fluid velocity is zero. The density ratio is $1000$, with $g=9.8$ m/s$^2$ and the properties of Table~\ref{tab:physical}.

The reference length is the head difference $H=0.36-0.018=0.342$ m, and $U_{\mathrm{ref}}=\sqrt{gH}=1.83074$ m/s. The comparisons use $144\times72\times24$ and $288\times144\times48$ cells, corresponding to $h=0.005$ and $0.0025$ m. Both momentum reconstructions are run to $t=1$ s from the same initial fields. The phase-density mean is harmonic unless the arithmetic control is explicitly identified. Surface tension and the same subgrid-scale closure are active in both reconstructions.

The wall-adapting local eddy-viscosity (WALE) model of Nicoud and Ducros \cite{NicoudDucros1999} uses the velocity-gradient tensor $g_{ab}=\partial_bu_a$, its symmetric part $S$, and the symmetric deviatoric part of $g^2$:

\begin{equation}\label{eq:wale_tensors}
S=\frac12(g+g^T),\qquad
S^d=\frac12\{g^2+(g^2)^T\}-\frac13 I\operatorname{tr}(g^2).
\end{equation}

\begin{equation}\label{eq:wale}
\nu_t=(C_w\Delta)^2
\frac{(S^d:S^d)^{3/2}}{(S:S)^{5/2}+(S^d:S^d)^{5/4}},
\qquad C_w=0.5,\quad\Delta=h.
\end{equation}

The viscosity is zero for a vanishing velocity gradient. Gradients are evaluated at faces using normal differences and centered transverse derivatives. The six face values adjacent to a cell are averaged to form $\mu_{t,i}=\rho_i\tfrac16\sum_{f\in\partial i}\nu_{t,f}$. The SGS stress uses Eq.~\eqref{eq:viscous_face} with $\mu$ replaced by $\mu_t$. Molecular and SGS work are accumulated separately. The model and coefficient are identical for Centered and WENO3-JS, while their realized SGS work depends on the resolved gradients.

\subsection{Positive phase-mass transport under large deformation}\label{sec:3d_positivity}

Arithmetic and harmonic intrinsic-density means are compared with identical centered momentum transport, pressure-increment relaxation, surface tension, and WALE closure. Both start from the same quiescent water column; Figure~\ref{fig:ar_h} compares the minimum gas partial densities during the stage updates. On $144\times72\times24$ cells, the arithmetic run develops negative gas partial density, detected at $t=0.922803$ s, while the harmonic run retains positive phase masses through $t=1$ s. The arithmetic calculation on $288\times144\times48$ cells also develops negative gas partial density, detected at $t=0.273883$ s. At the same resolution, both harmonic momentum reconstructions reach $t=1$ s with positive phase masses. The arithmetic detection times refer to the stage-level phase-mass diagnostics.

The difference follows from the mass carried by a given outgoing phase-volume flux. If the donor intrinsic density $r_{\ell,i}$ is much smaller than its neighbor, the arithmetic face density can be arbitrarily large relative to $r_{\ell,i}$. The harmonic density instead satisfies $r^H_{\ell,f}\le2r_{\ell,i}$. Consequently, the outgoing mass over a constituent Euler step is bounded by $2C_{\alpha_\ell,i}m_{\ell,i}$, and Eq.~\eqref{eq:positivity} provides a positive remaining mass when $C_{\alpha_\ell,i}<1/2$. The bound depends on the local phase mass and actual phase-volume flux, including regularization; it does not require the velocity divergence to vanish.

\paperfig{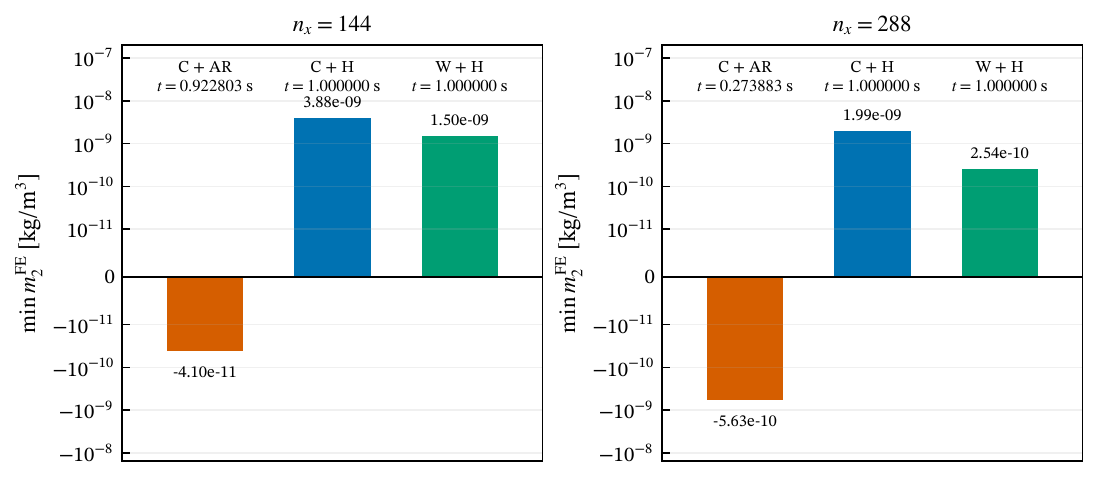}{Minimum gas partial density over the monitored constituent Euler stages. Arithmetic runs terminate after negative mass is detected at the indicated times; harmonic runs on both grids reach $t=1$ s. A symmetric logarithmic scale shows both signs. AR and H denote arithmetic and harmonic intrinsic-density means; C and W denote centered and WENO3-JS momentum transport.}{fig:ar_h}

\begin{table}[htbp]
\centering\small
\caption{Stage phase-mass diagnostics for the 3D dam break. The time column is the diagnostic endpoint. Phase partial densities are in kg/m$^3$; $n_x$ identifies the grids in Section~\ref{sec:3dsetup}.}\label{tab:3d_mass}
\begin{tabular}{@{}lrrrr@{}}\toprule
Method, $n_x$ & $t$ [s] & $\min m_1^{\mathrm{FE}}$ & $\min m_2^{\mathrm{FE}}$ & $\max C_\alpha$\\\midrule
Centered AR, 144 & 0.922803 & $8.848\times10^{-6}$ & $-4.096\times10^{-11}$ & 0.01259\\
Centered AR, 288 & 0.273883 & $9.838\times10^{-6}$ & $-5.627\times10^{-10}$ & 0.01172\\
Centered H, 144 & 1.000000 & $8.026\times10^{-6}$ & $3.879\times10^{-9}$ & 0.01259\\
WENO3-JS H, 144 & 1.000000 & $8.865\times10^{-6}$ & $1.501\times10^{-9}$ & 0.02969\\
Centered H, 288 & 1.000000 & $6.787\times10^{-6}$ & $1.989\times10^{-9}$ & 0.02275\\
WENO3-JS H, 288 & 1.000000 & $8.064\times10^{-6}$ & $2.545\times10^{-10}$ & 0.02980\\
\bottomrule
\end{tabular}
\end{table}

Both harmonic momentum reconstructions retain positive masses throughout the 144-grid calculation. Their largest outgoing phase-volume Courant numbers are $0.01259$ and $0.02969$, respectively, comfortably below the sufficient threshold. The maximum relative error of either integrated phase mass is below $1.1\times10^{-14}$. For the centered case, the intrinsic-density ratio $r_\ell/\rho_\ell^0$ in cells with $\alpha_\ell>1/2$ stays between $0.999596$ and $1.000305$. Thus the harmonic formulation handles the trace-phase transport while preserving global phase masses and small density variations in the bulk phases.

The arithmetic HIT calculation in Section~\ref{sec:hit_mean} also completes successfully. The benefit of harmonic averaging is most apparent in the demanding phase-mass outflow of the dam break. Harmonic averaging supplies a local outflow bound while preserving the same centered kinetic-energy identity.

\subsection{Interface evolution and energy pathways}\label{sec:3d_energy}

\paperfig{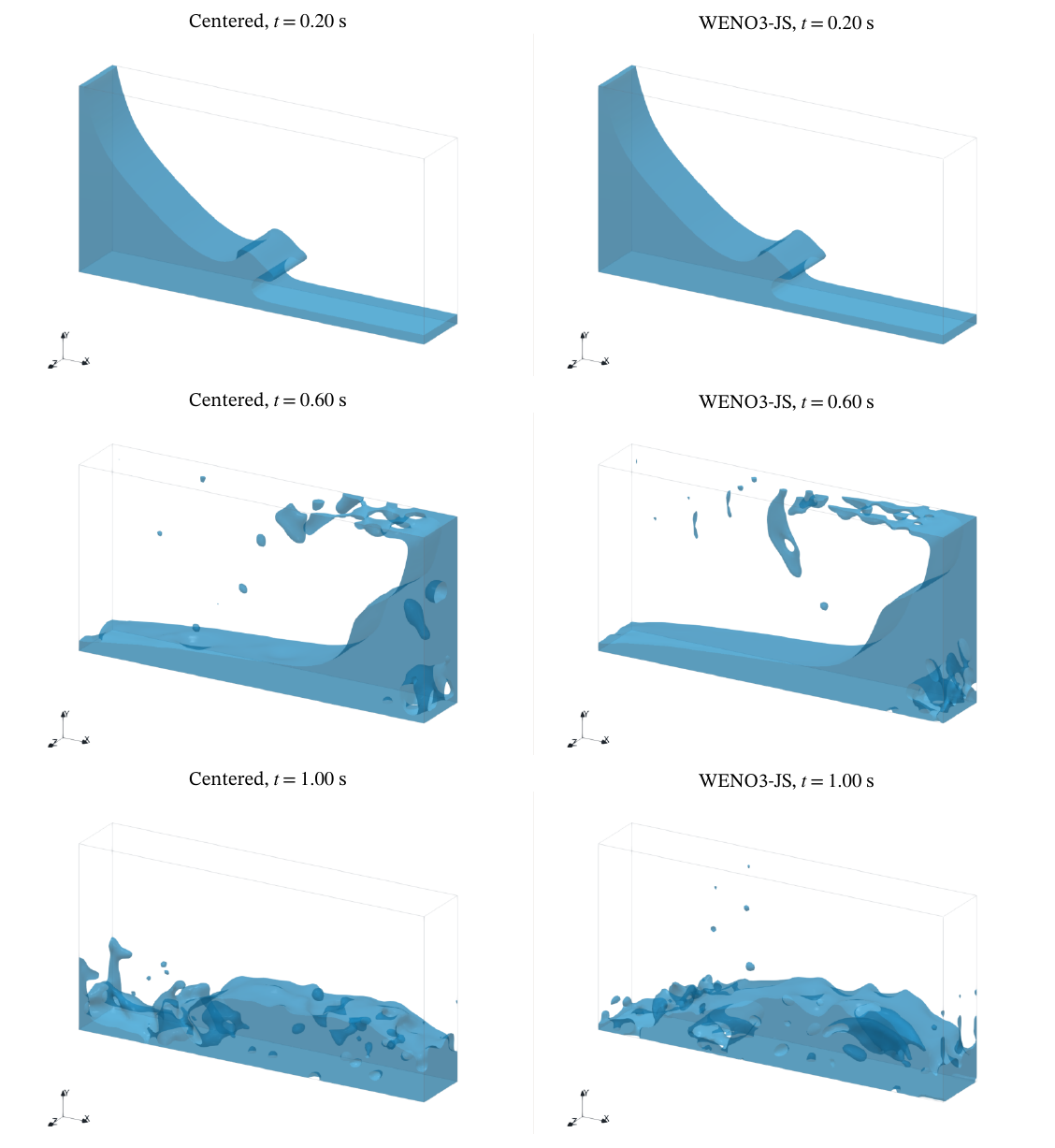}{Three-dimensional liquid region on $144\times72\times24$ cells at $t=0.20$, $0.60$, and $1.00$ s. Left: centered; right: WENO3-JS. Both use harmonic intrinsic-density averaging, pressure-increment relaxation, and WALE. The blue surface bounds $\phi\ge0.5$, including its contact with the tank walls.}{fig:3d_low}

Figure~\ref{fig:3d_low} follows the advancing surge, wall impact, and returning fragmented liquid. Both momentum reconstructions capture the large-scale sequence with the same physical parameters. Differences become visible in the interfacial folds and liquid fragments during the later motion. The energy budget below relates this behavior to the distinct transport and SGS contributions.

\paperfig{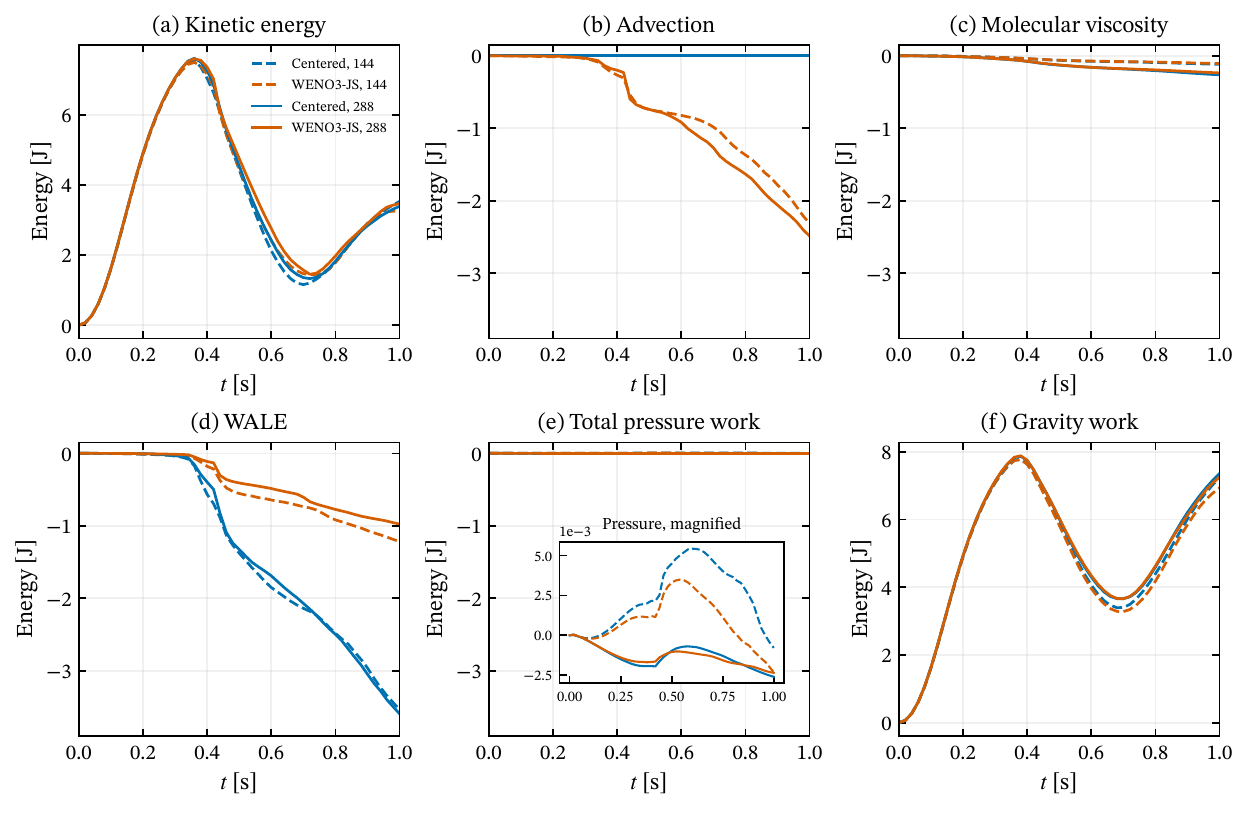}{Three-dimensional dam-break energy histories, in joules. Dashed and solid curves denote $144\times72\times24$ and $288\times144\times48$ cells, respectively. The cumulative advection, molecular-viscosity, WALE, and total-pressure panels use a common vertical scale. The pressure inset resolves the small total pressure work, including predictor work and the final increment correction.}{fig:3d_budget}

\begin{table}[htbp]
\centering\small
\caption{Cumulative energy budget at $t=1$ s on $144\times72\times24$ cells, in joules. $K(0)=0$. The total-pressure row is the sum of the two preceding pressure rows and is not an additional contribution.}\label{tab:3d_budget}
\begin{tabular}{@{}lrr@{}}\toprule
Contribution & Centered & WENO3-JS\\\midrule
$K(1)$ & 3.5298908 & 3.2614146\\
$W_{\mathrm{adv}}$ & $4.987\times10^{-20}$ & -2.3112683\\
$W_\mu$ & -0.1180166 & -0.1065419\\
$W_{\mathrm{SGS}}$ & -3.5384554 & -1.2169268\\
$W_g$ & 7.2449084 & 6.9404537\\
$W_{p,\mathrm{RK}}$ & -0.0006437 & -0.0021733\\
$\Delta K_{p,\mathrm{corr}}$ & -0.0001506 & -0.0001519\\
$W_{p,\mathrm{total}}$ & -0.0007943 & -0.0023253\\
$\Delta K_\sigma$ & -0.0577512 & -0.0419768\\
$R_{\mathrm{tr,time}}$ & $-1.357\times10^{-11}$ & $8.834\times10^{-11}$\\
$\varepsilon_K$ & $-3.067\times10^{-12}$ & $-2.587\times10^{-13}$\\
\bottomrule
\end{tabular}
\end{table}

Table~\ref{tab:3d_budget} gives final kinetic energies of $3.52989$ J and $3.26141$ J, but the mechanisms responsible for energy removal differ substantially. Centered advection contributes $4.99\times10^{-20}$ J, while WENO3-JS removes $2.31127$ J through advection. The corresponding WALE works are $-3.53846$ and $-1.21693$ J. With the centered method, WALE supplies $96.8\%$ of the combined molecular and SGS loss. With WENO3-JS, advection supplies $63.6\%$ of the combined advective, molecular, and SGS loss. The explicit model therefore becomes the principal dissipation mechanism when the upwind advective sink is removed.

The stronger SGS work is the response of the same WALE closure to the different resolved gradients. Centered transport keeps the energy transfer to that explicit operator visible in the budget. In the centered calculation, the total pressure contribution is $-7.9430\times10^{-4}$ J, only $0.0224\%$ of the WALE loss. Its predictor and correction components are both included. The absolute final budget closure error is below $3.1\times10^{-12}$ J for the two calculations, and the transport time residual is below $9\times10^{-11}$ J. The pressure-increment coupling thus preserves the low-dissipation transport advantage in this strongly deforming wall-bounded flow.

\subsection{Refined-grid comparison}\label{sec:3d_refined}

The $288\times144\times48$ calculations use the same physical parameters, harmonic intrinsic-density flux, pressure algorithm, and WALE coefficient. Refinement halves both the diffuse-interface thickness and the LES filter width in physical units, while preserving their ratios to the grid spacing. The arithmetic-mean calculation produces negative phase mass at this resolution (Table~\ref{tab:3d_mass}).

\paperfig{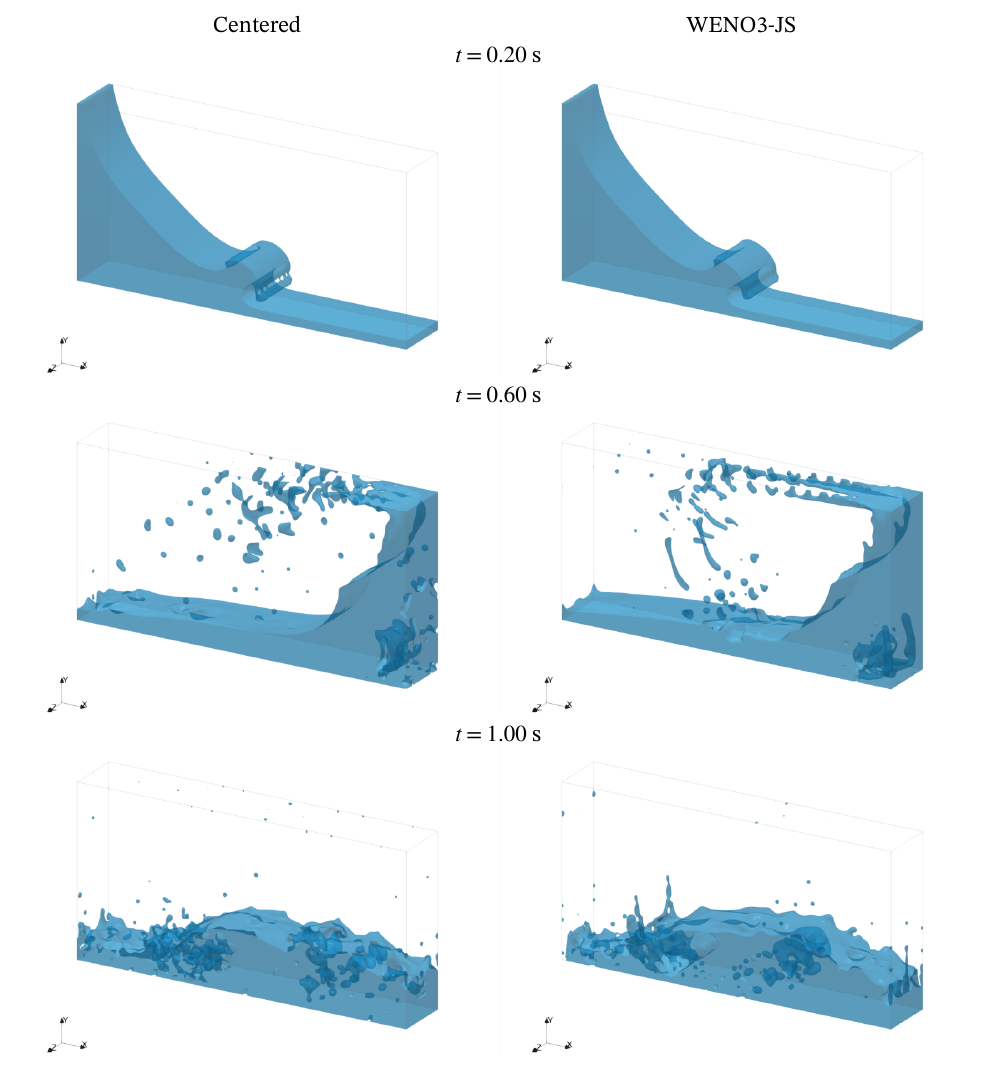}{Refined-grid liquid regions at $t=0.20$, $0.60$, and $1.00$ s. Results on $288\times144\times48$ cells: left, Centered; right, WENO3-JS. Both use harmonic intrinsic-density averaging and the same liquid-region definition and viewpoint as Fig.~\ref{fig:3d_low}.}{fig:3d_medium}

Figure~\ref{fig:3d_medium} shows the refined-grid interface evolution, including the finer liquid fragments resolved during impact and return flow. Both harmonic reconstructions retain positive phase masses through $t=1$ s. The largest outgoing phase-volume Courant numbers are $0.02275$ for Centered and $0.02980$ for WENO3-JS, satisfying the sufficient condition in Eq.~\eqref{eq:positivity}. The maximum relative error of either integrated phase mass is below $3.1\times10^{-14}$. For the centered calculation, the bulk intrinsic-density ratios remain between $0.999300$ and $1.000478$. Together with the arithmetic control, these results extend the harmonic flux's positive-mass transport to the finer grid without changing the centered momentum flux.

The energy decomposition in Fig.~\ref{fig:3d_budget} retains the same distinction between the two transport methods on the finer grid. Centered advective work is $-4.83\times10^{-19}$ J, while molecular viscosity and WALE remove $0.26282$ J and $3.59332$ J. WALE therefore accounts for $93.2\%$ of their combined loss. In the WENO3-JS calculation, advection removes $2.48768$ J, or $67.2\%$ of the combined advective, molecular, and SGS loss. The total centered pressure work is $-0.002606$ J, only $0.0725\%$ of the WALE loss. The final kinetic energies are $3.39100$ J and $3.46992$ J. The budget identifies the different dissipation mechanisms even when their final kinetic energies are close.

%% file: discussion.tex
\section{Discussion}\label{sec:discussion}

\subsection{Low-dissipation weakly compressible transport}\label{sec:contribution}

The proposed method combines centered transport with a local explicit pressure calculation for single- and two-phase flows. Its construction addresses two coupled requirements: retaining the kinetic energy carried by momentum advection and transporting phase mass robustly through strongly deformed interfaces. Symmetric velocity interpolation provides the semi-discrete advective cancellation, while harmonic intrinsic-density averaging bounds the mass carried by outgoing phase volume. The two properties coexist because the kinetic-energy identity uses the common mass flux without fixing its density interpolation.

The HIT comparisons quantify the resulting difference from upwind transport. At $t^*=8$ on $128^3$ cells, the magnitudes of the centered relative kinetic-energy changes are below $7.5\times10^{-8}$ in single-phase flow and $1.2\times10^{-7}$ in two-phase flow. WENO3-JS retains $3.60\%$ and $30.72\%$ of $K_0$, respectively. In the shear layers, the centered method retains stronger vortical structures while allowing the prescribed viscosity to dissipate energy. Rising bubbles and dam breaks demonstrate the same formulation in buoyant motion, capillary deformation, impact, and fragmentation.

\subsection{Pressure coupling and the resolved energy budget}\label{sec:pressure_discussion}

The pressure-increment formulation preserves the benefit of centered advection in the coupled calculation. Including $p^n$ in the transport stages allows the pressure gradient to act while the velocity field evolves. The subsequent EPP iterations correct the pressure through the change from that same $p^n$, and the final momentum correction applies this increment once. In the matched HIT comparison in \ref{app:coupling}, this temporal connection reduces the magnitude of the energy change by more than four orders relative to the full-pressure correction, despite identical centered spatial operators and acoustic settings.

The pressure contribution is measured as predictor pressure work plus the actual kinetic-energy change caused by the final correction. This decomposition explains the small energy changes of either sign in two-phase HIT as well as the small net loss in the viscous shear layer and dam break. Section~\ref{sec:pressure_property} establishes the conditional energy non-increase between successive pressure trials. Equations~\eqref{eq:pressure_energy}--\eqref{eq:full_budget} connect those trials to the applied momentum correction and predictor work, accounting for either sign of the measured total pressure contribution. For the centered $256^2$ two-phase shear layer, the magnitude of total pressure work is $0.0060\%$ of the molecular-viscous loss; for the 144- and 288-grid 3D dam breaks, it is $0.0224\%$ and $0.0725\%$ of the WALE loss, respectively. The pressure treatment therefore preserves the practical advantage of nondissipative advection in these cases.

\subsection{Implications for explicit SGS modeling and large-scale computation}\label{sec:les_implications}

The 3D energy budget shows how momentum reconstruction changes the role of an explicit SGS model. With the same WALE coefficient, centered transport leaves the principal dissipation to molecular and SGS stresses; WALE provides $96.8\%$ and $93.2\%$ of their combined loss on the 144 and 288 grids. In the WENO3-JS calculation, $63.6\%$ and $67.2\%$ of the combined advective, molecular, and SGS loss instead come from momentum advection on the two grids. Numerical damping also modifies the gradients on which the explicit model acts, accounting for the different SGS work. Removing that advective sink makes the model's contribution directly identifiable and provides a useful basis for comparing explicit closures in multiphase LES.

The collocated finite-volume structure also supports future integration with GPU AMR solvers. All momentum components share the same cell hierarchy, and the phase-mass and momentum fluxes are explicit face quantities suitable for conservative coarse--fine coupling. For $N$ cells, three transport stages and $N_p$ pressure iterations require $O[(3+N_p)N]$ work and $O(N)$ storage. Regular Cartesian stencils use neighboring halo exchanges, with scalar reductions for step-size estimates and diagnostics. The acoustic time-step restriction remains, but no global elliptic convergence process is required. These properties motivate extending the present low-dissipation formulation within the large-scale weakly compressible AMR framework demonstrated in our previous work \cite{Matsushita2021}.

\section{Conclusions}\label{sec:conclusions}

A centered finite-volume method with explicit pressure relaxation has been developed for weakly compressible single- and two-phase flows. The method combines kinetic-energy-preserving centered advection, finite-bound ACDI regularization, harmonic intrinsic-density transport, and a pressure-increment connection on a structured collocated grid. The shared mass flux and symmetric velocity interpolation cancel both semi-discrete advective production and dissipation of kinetic energy at finite velocity divergence. Harmonic averaging bounds phase-mass outflow and supplies a sufficient positivity condition for each constituent Euler stage without changing that transport identity.

The EPP construction includes the old pressure gradient in the transport predictor and corrects momentum once with the final pressure increment. Matching the cell pressure gradient to the divergence reconstructed from cell momentum gives a conditional energy non-increase estimate for the additional pressure iterations at fixed density and predictor. The pressure-increment connection retains the low dissipation of centered advection in the coupled algorithm. Under common pressure settings, the centered $128^3$ single- and two-phase HIT energies at $t^*=8$ differ from their initial values by $-7.47\times10^{-8}K_0$ and $-1.18\times10^{-7}K_0$, while WENO3-JS loses $96.40\%$ and $69.28\%$ of $K_0$. The comparison on $64^3$ cells supports the same separation between centered retention and upwind decay. The measured pressure work and transport-time residual explain these small remaining energy changes.

The shear-layer tests show the effect of low-dissipation transport on vortical motion, and the rising-bubble calculations reproduce the main rise and deformation trends of the incompressible benchmarks. In the density-ratio-1000 3D dam break, harmonic averaging maintains positive phase masses where the arithmetic control develops negative gas mass. The harmonic calculations conserve each integrated phase mass with relative errors below $1.1\times10^{-14}$ and $3.1\times10^{-14}$ on the 144 and 288 grids, respectively. With an identical WALE closure, the centered formulation removes the upwind advective energy sink and assigns the principal dissipation to the explicit stress model. Fully discrete budgets distinguish this contribution from molecular viscosity, pressure, capillarity, and time integration.

The method offers a low-dissipation route to large-scale multiphase LES with explicit SGS modeling. Its local pressure algorithm and collocated finite-volume form are designed with future GPU and AMR applications in view. Extending the demonstrated transport and pressure coupling to adaptive grids and complex boundaries is the next step toward practical large-scale flows.

%% file: appendices.tex
\appendix
\numberwithin{equation}{section}
\numberwithin{table}{section}
\numberwithin{figure}{section}
\renewcommand{\theequation}{\Alph{section}.\arabic{equation}}
\renewcommand{\thetable}{\Alph{section}.\arabic{table}}
\renewcommand{\thefigure}{\Alph{section}.\arabic{figure}}

\section{Pressure-relaxation energy estimate}\label{app:pressure}

This derivation concerns the additional pressure iterations in Eq.~\eqref{eq:p_relax}, with the predicted momentum and positive density fixed. Let $\delta\pi$ be an arbitrary change in the trial pressure increment. Expanding Eq.~\eqref{eq:pressure_functional} gives the exact finite difference

\begin{equation}\label{eq:pressure_variation}
\begin{aligned}
\mathcal K(\pi+\delta\pi)-\mathcal K(\pi)
&=-\Delta t\langle\boldsymbol u(p),G_c\delta\pi\rangle_V
+\frac{\Delta t^2}{2}\langle G_c\delta\pi,R^{-1}G_c\delta\pi\rangle_V\\
&=\langle\Delta tD_c\boldsymbol u(p),\delta\pi\rangle_V
+\frac12\langle\delta\pi,H_p\delta\pi\rangle_V,
\end{aligned}
\end{equation}

where $H_p=\Delta t^2G_c^\dagger R^{-1}G_c$ is the Hessian. The second line uses $G_c=-D_c^\dagger$ for the uniform periodic grid. Thus $g_p=\nabla_\pi\mathcal K=\Delta tD_c\boldsymbol u(p)$. The pressure update is $\pi^+=\pi-Bg_p$, with $B=\operatorname{diag}(\rho^*c^2)$ positive definite. Substituting $\delta\pi=-Bg_p$ and setting $z=B^{1/2}g_p$ yields

\begin{equation}\label{eq:pressure_proof}
\begin{aligned}
\mathcal K(\pi^+)-\mathcal K(\pi)
&=-\langle g_p,Bg_p\rangle_V
+\frac12\langle Bg_p,H_pBg_p\rangle_V\\
&=-\|z\|_V^2+\frac12\langle z,B^{1/2}H_pB^{1/2}z\rangle_V\\
&\le-\left[1-\frac12\lambda_{\max}(B^{1/2}H_pB^{1/2})\right]\|z\|_V^2.
\end{aligned}
\end{equation}

The linear term is nonpositive, while the quadratic term accounts for the finite pressure change. Equation~\eqref{eq:spectral_condition} makes their sum nonpositive by bounding the quadratic term. For constant density and acoustic speed, the eigenvalues are $(c\Delta t)^2\sum_d\sin^2(k_dh)/h^2$, displaying the acoustic time scale explicitly. Since $\sin^2(k_dh)\le1$, sufficient bounds for the additional pressure iterations are $c\Delta t/h<1$ in two dimensions and $c\Delta t/h<\sqrt{2/3}$ in three dimensions. These uniform-density bounds describe the pressure iterations alone. For variable density, the same estimate retains the actual density distribution through $R$ and $B$. Replacing $p$ by the increment $\pi=p-p^n$ leaves the Hessian unchanged. The resulting estimate therefore applies to the pressure-increment formulation used here, at each fixed predicted state; the energy change over a full physical step is given by Eq.~\eqref{eq:full_budget}.

\section{Dissipative form of the variable-viscosity operator}\label{app:viscous}

A periodic-grid identity clarifies the viscosity placement in Eq.~\eqref{eq:viscous_face}. Let $a_{ab}^{+}$ and $a_{ab}^{-}$ be the forward and backward differences of $u_a$ in direction $b$. Define the centered gradient and the difference of these one-sided slopes by

\begin{equation}\label{eq:visc_gradient}
g_{ab}=\frac{a_{ab}^{+}+a_{ab}^{-}}2,\qquad
e_{ab}=\frac{a_{ab}^{+}-a_{ab}^{-}}2,\qquad
S=\frac{g+g^T}{2},\quad S'=S-\frac13I\operatorname{tr}(g).
\end{equation}

Summation by parts with the face-averaged normal coefficient and the cell-weighted transverse term gives

\begin{equation}\label{eq:visc_dissipation}
\sum_iV_i\boldsymbol u_i\cdot(D\boldsymbol\tau)_i
=-\sum_iV_i\mu_i\left[
2S'_i:S'_i+\sum_{a\ne b}e_{ab,i}^2+\frac43\sum_a e_{aa,i}^2
\right]\le0.
\end{equation}

For the two-dimensional case, the tensors are embedded in three dimensions with zero out-of-plane components. Every term inside the brackets is nonnegative for $\mu_i\ge0$. The expression is also symmetric as a quadratic form, and it extends directly to a nonnegative frozen SGS viscosity. This identity explains why averaging $\mu D_a^cu_d$ as a product is retained in the face stress.

\section{Velocity reconstruction used in the upwind comparison}\label{app:weno}

For a positive total mass flux through face $i+1/2$, each velocity component $v$ is reconstructed from $v_{i-1},v_i,v_{i+1}$ as

\begin{equation}\label{eq:weno_reconstruction}
v^M_{i+1/2}=w_0\left(-\frac12v_{i-1}+\frac32v_i\right)
+w_1\left(\frac12v_i+\frac12v_{i+1}\right).
\end{equation}

\begin{equation}\label{eq:weno_weights}
\begin{aligned}
\beta_0&=(v_i-v_{i-1})^2,\qquad \beta_1=(v_{i+1}-v_i)^2,\\
\widehat w_0&=\frac{1/3}{(\epsilon_W+\beta_0)^2},\qquad
\widehat w_1=\frac{2/3}{(\epsilon_W+\beta_1)^2},\qquad
w_k=\frac{\widehat w_k}{\widehat w_0+\widehat w_1}.
\end{aligned}
\end{equation}

The negative-flux stencil is obtained by reflection. The regularization parameter is $\epsilon_W=10^{-6}$ in the squared-velocity units of each test. The reconstruction is applied componentwise only to momentum transport; the phase-volume and phase-mass fluxes remain those of Eq.~\eqref{eq:phase_fluxes}.

\section{Pressure-coupling comparison in inviscid turbulence}\label{app:coupling}

We next examine how the placement of the pressure force within a time step affects energy retention. In the comparison calculation, pressure is omitted from the momentum predictor and the full pressure is applied afterward as $-\Delta tG_cp^{n+1}$. In the proposed method, $-G_cp^n$ acts throughout the predictor and only the increment $-\Delta tG_c(p^{n+1}-p^n)$ is applied afterward. We refer to these as full-pressure and pressure-increment correction, respectively. Both calculations use the same centered spatial operators, initial fields, $c=640$, time-step ceiling $2\times10^{-5}$, and 20 pressure iterations. Single-phase and two-phase harmonic cases are compared through $t^*=2$.

\begin{table}[htbp]
\centering\small
\caption{Normalized kinetic energy $K/K_0$ for the pressure-coupling comparison at $t^*=2$ on $64^3$ cells. The acoustic speed, time step, and centered spatial operators are common.}\label{tab:pressure_coupling}
\begin{tabular}{@{}lrr@{}}\toprule
Flow & Full-pressure correction & Pressure-increment correction\\\midrule
Single phase & 0.9990517375 & 0.9999999292\\
Two phases, H & 0.9964133792 & 1.0000001286\\
\bottomrule
\end{tabular}
\end{table}

Table~\ref{tab:pressure_coupling} shows that the full-pressure correction loses $9.48\times10^{-4}K_0$ in single-phase flow and $3.59\times10^{-3}K_0$ in two-phase flow. With the pressure-increment connection, the corresponding energy changes are $-7.08\times10^{-8}K_0$ and $+1.29\times10^{-7}K_0$. Thus the temporal pressure connection reduces the magnitude of the energy change by more than four orders in these matched tests. The spatial advection remains centered in both variants, demonstrating that retaining low dissipation requires attention to the pressure coupling as well as the momentum flux.

\section{Parasitic currents and energy balance in a static drop}\label{app:static_drop}

Parasitic currents can arise from an imbalance between discrete pressure and capillary forces even when momentum advection itself neither produces nor dissipates kinetic energy. A static circular drop at density ratio $1000$ is used to measure this residual motion and assess the effect of harmonic intrinsic-density interpolation relative to arithmetic interpolation. A matched inviscid calculation separates the residual energy input from its viscous attenuation, while the pressure jump and contour deformation quantify the effect on the resolved drop.

The liquid region has radius $R=0.1$ m and is centered in a $1\times1$ m square with no-slip walls. The reference densities are $(\rho_1^0,\rho_2^0)=(1000,1)$ kg/m$^3$, the dynamic viscosities are $(\mu_1,\mu_2)=(10^{-3},1.8\times10^{-5})$ Pa s, and $\sigma=0.0728$ N/m. Gravity and SGS stress are absent. With $r$ denoting distance from the drop center, the initial fields are $\chi=\{1+\tanh[(R-r)/(2\epsilon)]\}/2$, $\phi=\delta+(1-2\delta)\chi$, $m_\ell=\rho_\ell^0\alpha_\ell$, $\boldsymbol u=0$, and $p=10^5+(\sigma/R)\chi$ Pa. The initial and transported widths satisfy $\epsilon/h=0.51$. This diffuse Laplace-pressure profile is evolved with the numerical curvature and capillary-force operators of Section~\ref{sec:forces}.

For $D=2R$, define $\tau_\sigma=(\rho_1^0D^3/\sigma)^{1/2}=10.483$ s, $U_\sigma=D/\tau_\sigma$, and $E_{\sigma,0}=2\pi R\sigma$. The initial circular-interface energy $E_{\sigma,0}$ normalizes the kinetic energy per unit depth, since $K(0)=0$. Centered harmonic calculations on $32^2$, $64^2$, and $128^2$ cells extend to $t/\tau_\sigma=10$. They use $c=10$ m/s, $\Gamma=1.0192U_{\max}$, $(C_a,C_d,C_\sigma)=(0.2,0.08,0.1)$ in Eq.~\eqref{eq:dt}, and respective time-step ceilings of $5\times10^{-4}$, $2.5\times10^{-4}$, and $1.25\times10^{-4}$ s. Three controls on $64^2$ cells separately replace the harmonic mean by the arithmetic mean, set both viscosities to zero, or halve the time-step ceiling, retaining the other settings.

\paperfig{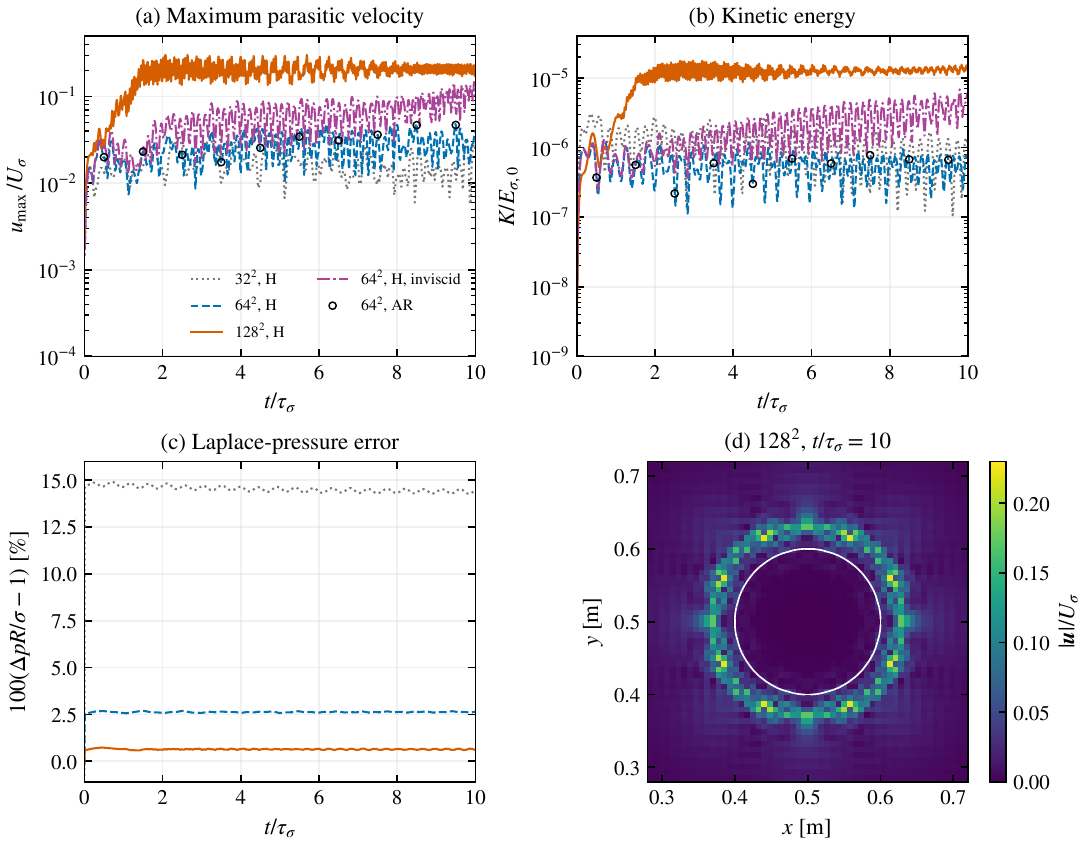}{Static drop at density ratio $1000$. (a) Maximum parasitic speed $u_{\max}=\max_i|\boldsymbol u_i|$; (b) kinetic energy; (c) relative pressure-jump error, using mean pressures in $\chi>0.99$ and $\chi<0.01$; (d) velocity magnitude on $128^2$ cells at $t/\tau_\sigma=10$, with the $\phi=0.5$ contour in white. H and AR denote harmonic and arithmetic intrinsic-density means. All cases use centered momentum transport. The arithmetic-control symbols overlap the $64^2$ harmonic histories. The inviscid control has zero viscosity in both phases. Panel (c) compares the three viscous harmonic grids.}{fig:static_drop_assessment}

The matched $64^2$ harmonic and arithmetic calculations give indistinguishable histories at the plotted scale (Fig.~\ref{fig:static_drop_assessment}a,b); their maximum difference in $K/E_{\sigma,0}$ is $1.34\times10^{-15}$. Thus the change of intrinsic-density mean does not measurably alter the parasitic motion in this test. Refinement reduces the final pressure-jump error from $14.47\%$ to $2.65\%$ and $0.603\%$ (Fig.~\ref{fig:static_drop_assessment}c). This error includes the finite interface width as well as the curvature discretization. The peak $u_{\max}/U_\sigma$, however, is $0.0352$, $0.0488$, and $0.306$, respectively; maximum parasitic speed does not decrease with refinement.

On $128^2$ cells, the strongest motion lies outside the liquid contour, whereas the liquid region remains nearly circular (Fig.~\ref{fig:static_drop_assessment}d). Over the observation interval, the maximum root-mean-square radial deviation from the equal-area circle is $0.138\%$ of $R$, the contour area changes by at most $0.063\%$ from its initial value, and $K/E_{\sigma,0}<1.79\times10^{-5}$. The remaining motion therefore has a small effect on the resolved drop shape, despite its larger local velocity near the interface.

The fully discrete budget explains how this residual motion is sustained (Table~\ref{tab:static_budget}). Pressure work in the predictor and the capillary-stage energy change are individually much larger than their sum. Their signs depend on the update sequence, so the combined pressure--capillary contribution is considered. On $128^2$ cells, $(W_{p,\mathrm{total}}+\Delta K_\sigma)/E_{\sigma,0}=2.72\times10^{-4}$; molecular viscosity removes $2.57\times10^{-4}$, leaving $K/E_{\sigma,0}=1.49\times10^{-5}$ at $t/\tau_\sigma=10$. The cumulative viscous loss is $17.3$ times the final kinetic energy, while net advective work is negligible. The small remaining energy thus reflects viscous removal of a finite residual forcing.

\begin{table}[htbp]
\centering\small
\caption{Static-drop energies and cumulative contributions in Eq.~\eqref{eq:full_budget} at $t/\tau_\sigma=10$, normalized by $E_{\sigma,0}$. All columns use centered advection and harmonic intrinsic-density averaging. The final row reports the maximum absolute closure residual over the observation interval.}\label{tab:static_budget}
\begin{tabular}{@{}lrrr@{}}\toprule
Contribution & $64^2$ & $128^2$ & $64^2$, inviscid\\\midrule
$K$ & $6.827\times10^{-7}$ & $1.485\times10^{-5}$ & $5.643\times10^{-6}$\\
$W_{\mathrm{adv}}$ & $8.678\times10^{-26}$ & $-2.553\times10^{-24}$ & $-1.957\times10^{-25}$\\
$W_\mu$ & $-2.852\times10^{-6}$ & $-2.567\times10^{-4}$ & $0$\\
$W_{p,\mathrm{RK}}$ & $5.5937777\times10^{-3}$ & $5.7037464\times10^{-3}$ & $5.5873527\times10^{-3}$\\
$\Delta K_{p,\mathrm{corr}}$ & $-4.960\times10^{-10}$ & $-7.711\times10^{-11}$ & $-4.963\times10^{-10}$\\
$\Delta K_\sigma$ & $-5.5902419\times10^{-3}$ & $-5.4321784\times10^{-3}$ & $-5.5817094\times10^{-3}$\\
$R_{\mathrm{tr,time}}$ & $2.177\times10^{-15}$ & $3.420\times10^{-15}$ & $-1.891\times10^{-15}$\\
$\max_t|\varepsilon_K|$ & $1.633\times10^{-15}$ & $1.394\times10^{-13}$ & $8.262\times10^{-15}$\\
\bottomrule
\end{tabular}
\end{table}

Halving the $64^2$ time-step ceiling changes the mean $K$ and mean $u_{\max}$ over $5\le t/\tau_\sigma\le10$ by $-0.142\%$ and $-0.0472\%$, respectively. All six calculations retain positive phase masses at every constituent Euler stage, with relative integrated phase-mass errors below $8.1\times10^{-14}$.

Without viscosity on $64^2$ cells, the positive net pressure--capillary contribution increases $K/E_{\sigma,0}$ from $2.49\times10^{-6}$ at $t/\tau_\sigma=5$ to $5.64\times10^{-6}$ at $10$. The final pressure-increment correction is a small negative contribution, and net advective work remains negligible. Thus the advective identity remains satisfied, but it does not prevent kinetic energy from being supplied by the residual pressure--capillary imbalance. This residual motion limits the accuracy of the present force discretization for nearly static capillary flows.

%% file: backmatter.tex
\section*{CRediT authorship contribution statement}
\textbf{Shintaro Matsushita:} Conceptualization, Data curation, Formal analysis, Funding acquisition, Investigation, Methodology, Software, Validation, Visualization, Writing -- original draft, Writing -- review \& editing.
\textbf{Suhas S. Jain:} Formal analysis, Methodology, Supervision, Writing -- review \& editing.
\textbf{Andrei G. Fedorov:} Funding acquisition, Project administration, Resources, Validation, Writing -- review \& editing.
\textbf{Tetsuya Suekane:} Funding acquisition, Supervision, Writing -- review \& editing.

\section*{Declaration of competing interest}
The authors declare that they have no known competing financial interests or personal relationships that could have appeared to influence the work reported in this paper.

\section*{Data availability}
The numerical data underlying the reported curves and energy-budget tables, together with case settings and post-processing instructions, will be available in Mendeley Data at \url{https://doi.org/10.17632/sdb8kgc3pt.1}. Full time-resolved three-dimensional volume fields are not included in the dataset because of their storage requirements.

\section*{Acknowledgements}
This work was supported by the Japan Society for the Promotion of Science (JSPS) KAKENHI Grant Numbers 22H03770, 22K14178, and 25K07582, and by the Japan Science and Technology Agency (JST) through Adopting Sustainable Partnerships for Innovative Research Ecosystem (ASPIRE), Grant Number JPMJAP2407. This study was carried out using the TSUBAME4.0 supercomputer at Institute of Science Tokyo. The funding agencies had no role in the study design; collection, analysis, or interpretation of data; preparation of the manuscript; or decision to submit the work for publication.

\section*{Declaration of generative AI and AI-assisted technologies in the manuscript preparation process}
During the preparation of this work, the authors used OpenAI ChatGPT and Codex to assist with English-language drafting and editing and with the analysis and visualization of numerical results. After using these tools, the authors reviewed and edited the content as needed and take full responsibility for the content of the published article.